\documentclass[a4paper,12pt]{article}
\usepackage{epsfig}
\usepackage{graphicx}
\usepackage{amsmath}
\usepackage{amsfonts}
\usepackage{amssymb,cite}
\newif\ifiTheta
\iThetatrue    
\newcommand{\ft}[2]{{\textstyle\frac{#1}{#2}}}

\def\trace{\mathop{\rm Tr}\nolimits}
\def\rmi{{\rm i}}
\def\rmd{{\rm d}}
\def\rme{{\rm e}}
\newsavebox{\uuunit}
\sbox{\uuunit}
    {\setlength{\unitlength}{0.825em}
     \begin{picture}(0.6,0.7)
        \thinlines
        \put(0,0){\line(1,0){0.5}}
        \put(0.15,0){\line(0,1){0.7}}
        \put(0.35,0){\line(0,1){0.8}}
       \multiput(0.3,0.8)(-0.04,-0.02){12}{\rule{0.5pt}{0.5pt}}
     \end {picture}}
\newcommand {\unity}{\mathord{\!\usebox{\uuunit}}}

\csname @addtoreset\endcsname{equation}{section}
\newcommand{\upomega}{\mbox{\usefont{U}{psy}{m}{n}w}}

\newcommand{\cG}{{\cal G}}
\newcommand{\cH}{{\cal H}}
\newcommand{\cL}{{\cal L}}
\newcommand{\cM}{{\cal M}}
\newcommand{\cN}{{\cal N}}

\newcommand{\tI}{{\tilde I}}
\newcommand{\tJ}{{\tilde J}}

\newcommand{\la}{\lambda}
\newcommand{\si}{\sigma}

\newcommand{\bea}{\begin{eqnarray}}
\newcommand{\eea}{\end{eqnarray}}
\newcommand{\nnu}{\nonumber}
\newcommand{\be}{\begin{equation}}
\newcommand{\ee}{\end{equation}}
\newcommand{\bt}{\begin{tabular}}
\newcommand{\et}{\end{tabular}}
\newcommand{\ba}{\begin{array}}
\newcommand{\ea}{\end{array}}
\newcommand{\bbm}{\begin{bmatrix}}
\newcommand{\ebm}{\end{bmatrix}}

\newcommand{\prt}{\partial}

\newcommand{\prtsl}{\partial\negthickspace\negthickspace /}

\newcommand{\gs}{g_{\scriptscriptstyle S}}
\newcommand{\ga}{g_{\scriptscriptstyle A}}

\begin{document}

\begin{titlepage}
\begin{flushright}
SU-ITP-05/003\\
hep-th/0412081
\end{flushright}
\vspace{.5cm}
\begin{center}
\baselineskip=16pt {\bf \LARGE  $\mathcal{N}=4$ ``Fake'' Supergravity
}\\
\vfill
{\large
Marco Zagermann
} \\
\vfill \vspace{-2cm}
{\small
Department of Physics, Stanford University,\\
Varian Building, Stanford, CA 94305-4060, USA.\\ \vspace{6pt}
 }
\end{center}
\vfill
\begin{center}
{\bf Abstract}
\end{center}
{\small We study curved and flat
BPS-domain walls in 5D, $\mathcal{N}=4$ gauged supergravity
and show that their effective dynamics along the flow is described by a generalized form of  ``fake supergravity''.
This
 generalizes previous work in $\mathcal{N}=2$ supergravity 
and might hint towards a universal behavior of   gauged supergravity theories
in supersymmetric   domain wall backgrounds. 
We show that  BPS-domain walls          in 5D, $\mathcal{N}=4$
supergravity can never be curved if they are supported by the
supergravity scalar only. Furthermore, a purely Abelian gauge group
or a purely semisimple gauge group can never lead to a curved
domain wall, and the flat walls for these gaugings  always     exhibit a 
runaway behavior.

 }\vspace{2mm} \vfill \hrule width 3.cm
{\footnotesize \noindent e-mail: zagermann@itp.stanford.edu    }
\end{titlepage}
\tableofcontents{}

\baselineskip 6 mm

\section{Introduction}
Domain wall solutions of $(d+1)$-dimensional
supergravity theories have received a lot of
attention during the past few    years. This interest was largely
driven by applications in the context of holographic renormalization
group flows and  certain brane world models. 
In most of these applications, the domain walls of
interest preserve a fraction of the original  supersymmetry of 
the   supergravity theory
they are embedded in. A  domain wall of this type  can be either
Minkowski-sliced or AdS-sliced,
\begin{equation}
\rmd s^2=\rme^{2U(r)}g_{mn}(x)\,\rmd x^m \,\rmd x^n + \rmd r^2,
\label{curvedmetric0}
\end{equation}
depending on whether, respectively,   $g_{mn}$ is the metric of  $d$-dimensional
Minkowski- or Anti-de Sitter space. A  non-trivial   warp factor
$U(r)$ (i.e., one that does not give rise to $(d+1)$-dimensional Minkowski- or Anti-de Sitter space) 
requires a nontrivial scalar profile
 $\phi^{x}(r)$ $(x=1,\ldots, m)$, as dictated by
 the Einstein equations. A domain wall thus  defines a
 curve $\phi^{x}(r)$ on the scalar manifold.

The allowed scalar manifolds in supergravity theories are in general
highly constrained and  strongly depend upon the spacetime
dimension, the amount of supersymmetry,   as well as    on the type of
multiplet the scalars are sitting in. The geometrical constraints on
the scalar manifolds  also leave their trace in the BPS-equations of
the scalar fields, which are likewise  highly spacetime-,
supersymmetry- and multiplet dependent.

It came therefore as quite a surprise when it was found in
\cite{CCDVZ}  that one can reformulate the BPS-conditions for domain walls 
  in 5D,
$\mathcal{N}=2$ supergravity  in such a way that their naive
strong multiplet dependence effectively disappears. The same is true
for the scalar potential, which, in this simplified reformulation,
also contains the scalar fields from  vector and 
 hypermultiplets   in a symmetric way. In order to achieve this
simplification, one has to restrict one's attention to the effective
dynamics \emph{along the curve} $\phi^{x}(r)$  of a given BPS-domain
wall   and properly ``integrate out'' the orthogonal scalar fields.
Interestingly,   this also exactly reproduces the equations of
``fake supergravity'' that were  introduced in ref. \cite{Freedman:2003ax} to
prove the stability of domain walls in certain scalar/gravity
theories that, despite some superficial similarities, are not
necessarily supersymmetric\footnote{Some related work appeared in 
  \cite{Bak:2004yf}.}. The fake supergravity formalism in
\cite{Freedman:2003ax} was tailor-made to   describe       
           curved domain walls 
and   generalizes and refines
 the earlier work \cite{Townsend:1984iu,Skenderis:1999mm,DeWolfe:1999cp}.
It  was worked out in \cite{Freedman:2003ax}  in  detail  
 for theories with only one scalar field,
and it is this scalar field that one has to  identify with the
scalar direction along the flow curve $\phi^{x}(r)$ in    5D,
$\mathcal{N}=2$ supergravity. The fake supergravity equations were also 
generalized to several scalar fields in \cite{Freedman:2003ax}, but only for a
very particular type of scalar potential. One of the lessons of
\cite{CCDVZ}, however, is that a      generalization and
covariantization to more than one scalar field can go along various
different lines, and it seems that only the effective one-scalar
field formulation is universal.

The results of \cite{CCDVZ} are by no means of only formal interest.
On the contrary, it was found that the simplified reformulation of
true supergravity \`{a} la fake supergravity provides  a very handy
tool for studying   true BPS-domain walls themselves. For example,
using the simplified language of ``fake'' supergravity, 
it is  fairly  easy to prove that BPS-domain walls that are only
supported by scalars from   vector multiplets  can at most be
\emph{Minkowski-sliced}. An AdS-sliced BPS-domain wall thus must
involve non-trivial hypermultiplet scalars. This fact had gone
unnoticed before.

In this paper, we will go one step beyond the work of \cite{CCDVZ} and
study domain walls in             5D, $\mathcal{N}=4$ supergravity
along similar lines. That is, we will try to similarly recast the
BPS-equations and the scalar potential in a generalized ``fake''
supergravity form. This generalization is highly nontrivial due to
the following reasons:
\begin{itemize}
\item The BPS-constraints are stronger, as there are now twice as
many supersymmetries to preserve.
\item The $\mathcal{N}=4$ theory is
           $Usp(4) \cong SO(5)$   instead of $Usp(2)\cong SU(2)$
           covariant,
           i.e., several peculiarities of the group $SU(2)$ no
           longer hold.
\item The scalar manifolds in the $\mathcal{N}=4$ theory  are 
   of the type $SO(1,1)\times~SO(5,n)/(SO(5)~\times SO(n))$,
which  are,  in   general,   neither  very special nor   quaternionic manifolds.
Contrary to what happens in rigid supersymmetry, the $\mathcal{N}=4$
 theory can therefore not be viewed as a special case of the
$\mathcal{N}=2$ theory.
\end{itemize}
Given these differences,
 it is all the more intriguing that only few features of the
 $\mathcal{N}=2$ formulation are identified as
  $SU(2)$  artifacts and that one finds an exactly analogous picture: The
effective BPS-equations  and the scalar potential
can again be brought to a simple,
generalized ``fake'' supergravity-type form, no matter whether the
running scalar field sits in the $\mathcal{N}=4$ supergravity
multiplet or in an $\mathcal{N}=4$ vector or tensor multiplet. Just as 
in the $\mathcal{N}=2$ analogue \cite{CCDVZ}, we  can also use this simplified
language in order to study the domain walls themselves. It    is found 
that BPS-domain walls that are supported by the supergravity scalar
only are necessarily flat. Similarly, if the gauge group is purely
Abelian or purely semisimple, the domain wall can at most be flat,
no matter by which type of scalar field they are supported. Any
flat domain wall for these gaugings, however, has a runaway
behaviour. These results could prove very useful for studies of
holographic renormalization group flows  \cite{holflow}  
in the setup of, e.g.,    \cite{CGWZ}
   or for 
domain walls in
 gauged supergravities   that derive from flux compactifications (see e.g.,
\cite{Mayer:2004sd,House:2004hv,LouisVaula} for some   recent work in this direction). Moreover, the present  
work suggests that the language of ``fake supergravity''
is far more universal than previously thought and that it
 might well be applicable to
 a much wider range of gauged supergravity theories, perhaps, if properly  
formulated,   even to all of them. Fake supergravity might thus turn out to be
not that fake after all!   

The organization of this paper is as follows: In section 2, we
briefly recapitulate the structure of  BPS-domain walls in
                 5D, $\mathcal{N}=2$ gauged supergravity and the
                 relation to the fake supergravity formalism
                 developed in \cite{Freedman:2003ax}. In section 3, we then
                 discuss the structure of       5D, $\mathcal{N}=4$
                 gauged and ungauged supergravity and study its
                 $1/2$-supersymmetric domain wall solutions.
                 This is done by rewriting  the    BPS-equations and
                 the scalar potential in a generalized,
                                      ``$\mathcal{N}=4$'' fake
                                      supergravity form.
In this simplified version, several general statements about
possible                                  BPS-domain walls are
easily derived. We end with some conclusions in section 4.
Appendix A proves the equivalence of two flatness conditions.

\section{True and fake $5D$, $\mathcal{N}=2$ supergravity}
In this section, we briefly summarize the key results of \cite{CCDVZ}
 on  BPS-domain walls in
  true and fake,   5D, $\mathcal{N}=2$ supergravity.
For earlier work on (smooth) 
 flat and curved  BPS-domain walls in these
theories, see \cite{                  Behrndt:1999kz,             Kallosh:2000tj, Behrndt:2000tr, Gibbons:2000hg   
,Behrndt:2000km,Ceresole:2001wi,
Alekseevsky:2001if, Behrndt:2001km,Anguelova,Cacciatori:2003kv,Celi:2003qk} and
\cite{LopesCardoso:2001rt,Chamseddine:2001hx,Cardoso:2002ec,Behrndt:2002ee,Cardoso:2002ff},
respectively.

\subsection{5D, $\mathcal{N}=2$ gauged supergravity}
Five-dimensional, $\mathcal{N}=2$ supergravity can be coupled to
vector-, tensor- and hypermultiplets. The precise form of these
theories was derived in the original references
\cite{Gunaydin:1984bi,Gunaydin:1985ak,Sierra:1985ax,Gunaydin:1999zx,Ceresole:2000jd,
   Gunaydin:2000ph                         ,Bergshoeff:2004kh},
to which we refer the reader for further details.
 As was emphasized already in  \cite{Ceresole:2001wi,Ceresole:2001zf}, all the terms
in the theory that are due to the presence of tensor multiplets have to vanish on a BPS-domain wall background,
  and we   can thus restrict ourselves to
the coupling of $n_V$  vector multiplets and $n_H$
 hypermultiplets to supergravity.

The bosonic
 field content of such a theory consists of the f\"unfbein $e_\mu^m$,
$(n_V+1)$ vector fields $A_{\mu}^I$ $(I=0,1,\ldots,n_{V})$ and $(n_{V}+n_H)$
real scalar fields $(\varphi^{x},q^{X})$, with $x=1,\ldots,n_V$ and $X=1,\ldots, 4n_H$. Here, we have combined  the graviphoton of the supergravity
multiplet with the $n_V$ vector fields of the $n_V$ vector multiplets
to form a single $(n_V +1)$-plet $A_{\mu}^{I}$.


The $n_V$ scalar fields $\varphi^x$  of the vector multiplets
parameterize a ``very special'' real  manifold $\mathcal{M}_{\rm
VS}$, i.e., an     $n_V$-dimensional hypersurface of an auxiliary
$(n_V+1)$-dimensional   space spanned by coordinates $h^{I}$
$(I=0,1,\ldots, n_{V})$ :
\begin{equation}
\mathcal{M}_{\rm VS}=\{ h^{I}\in \mathbb{R}^{(n_V+1)}:
C_{IJK}h^I h^J h^K =1\}, \label{defVS}
\end{equation}
where the constants $C_{IJK}$ appear in a Chern-Simons-type coupling
 of the Lagrangian.
On $\mathcal{M}_{\rm VS}$, the $h^{I}$ become functions of the $n_V$
physical scalar fields, $\varphi^{x}$. The metric, $g_{xy}$,  on the
very special manifold is determined via
\begin{equation}
g_{xy}=-3 C_{IJK} (\partial_{x}h^I)(\partial_{y}h^{J}) h^{K}.
\end{equation}
%

The scalars  $q^X$ $(X=1,\ldots, 4n_{H})$ of  $n_{H}$
hypermultiplets, on the other hand, take their values in a
quaternionic-K{\"a}hler manifold $\mathcal{M}_{\rm Q}$
\cite{Bagger:1983tt}, i.e., a manifold of real dimension $4n_H$ with
holonomy group contained in $SU(2)\times USp(2n_H)$. The vielbein on
this manifold is denoted   by $f_X^{iA}$, where $i=1,2$,  and
$A=1,\ldots,2n_{H}$ refer to an adapted $SU(2)\times USp(2n_{H})$
decomposition of the tangent space. The hypercomplex structure is
$(-2)$ times the curvature of the $SU(2)$ part of the holonomy
group, denoted as $\mathcal{R}^{rZX}$ $(r=1,2,3)$, so that the
quaternionic identity reads
\begin{equation}
  {\cal R}^r_{XY}{\cal R}^{sYZ}=-\ft14\,\delta ^{rs}\,\delta _X{}^Z
   -\ft12\,\varepsilon ^{rst}\,{\cal R}^t_X{}^Z.
 \label{quaterid}
\end{equation}
The vector fields $A_{\mu}^I$ can be used to gauge up to $(n_V+1)$
isometries of the quaternionic manifold $\mathcal{M}_Q$ (provided
such isometries exist)\footnote{A non-Abelian gauge group also has
to leave the $C_{IJK}$ invariant, which implies that the gauge group
also has to  be a subgroup of the isometry group of $\mathcal{M}_{VS}$
\cite{Gunaydin:1985ak,Gunaydin:1999zx,Ellis:2001xd,Bergshoeff:2004kh}.}.

The quaternionic Killing vectors, $K_{I}^{X}(q)$, that generate these
isometries on $\mathcal{M}_{\rm Q}$ can be expressed in terms of the
derivatives of $SU(2)$ triplets of Killing prepotentials (or
``moment maps'') $P_{I}^{r}(q)$ $(r=1,2,3)$ via
\begin{equation}
D_{X}P^{r}_{I}= \mathcal{R}_{XY}^{r}K^{Y}_{I}, \qquad \Leftrightarrow
\qquad \left\{ \begin{array}{c}
  K_{I}^Y=-\frac{4}{3}\mathcal{R}^{rYX}D_{X}P^{r}_{I} \\ [2mm]
  D_XP_I^r=-\varepsilon ^{rst}{\cal R}^s_{XY}D^YP_I^t,
\end{array}\right.
\label{KillingP}
\end{equation}
where $D_{X}$ denotes the $SU(2)$ covariant derivative, which
contains the $SU(2)$ connection $\omega_{X}^{r}$ with curvature $
\mathcal{R}^r_{XY}$:
\begin{equation}
  D_X P^r= \partial _XP^r+2\,\varepsilon ^{rst} \omega _X^sP^t,\qquad
\mathcal{R}^r_{XY}=2\,\partial _{[X}\omega _{Y]}^r+2\,\varepsilon
^{rst}\omega _X^s\omega _Y^t.
 \label{defSU2cR}
\end{equation}
The prepotentials  have to  satisfy the constraint
\begin{equation}\label{constraint}
\frac{1}{2}\mathcal{R}_{XY}^{r}K_{I}^{X}K_{J}^{Y}-
\varepsilon^{rst}P_{I}^{s}P_{J}^{t} +\frac{1}{2}f_{IJ}{}^{K}P_{K}^{r}=0,
\end{equation}
where $f_{IJ}{}^{K}$ are the structure constants of the gauge group.
In this section, we will frequently switch between the above vector
notation for $\mathfrak{su}(2)$-valued quantities such as
$P_{I}^{r}$, and the usual
   $(2\times 2)$   matrix notation,
\begin{equation}
\mathbf{P}_I= \left(   P_{Ii}{}^j\right) ,\qquad
P_{Ii}{}^j\equiv \rmi\,\sigma_{ri}{}^jP_{I}^{r}  ,
 \label{matrixVector}
\end{equation}
where boldface expressions such as $\mathbf{P}_I$  refer to the
$(2\times 2)$-matrices with the indices $i,j$ suppressed.
Turning on only the metric and the scalars, the Lagrangian of
such a gauged supergravity  theory
is
\begin{equation}
e^{-1}\mathcal{L}=-\frac{1}{2}R-\frac{1}{2}g_{xy}\partial_{\mu}\varphi^{x}
\partial^{\mu}\varphi^{y}-\frac{1}{2}g_{XY}\partial_{\mu}q^{X}
\partial^{\mu}q^{Y}-g^2\mathcal{V}(\varphi,q),
\end{equation}
whereas the supersymmetry transformation laws of the fermions are given by
\begin{eqnarray}
\delta \psi_{\mu i}&=& {\nabla}_\mu\epsilon_i-  \omega_{\mu i}{}^{j}
\epsilon_{j}- \frac{\rmi}{\sqrt{6}}\,
g\,\gamma_{\mu}P_{i}^{\,\, j}\epsilon_{j},\label{gravitino}\\
\delta \lambda_{i}^{x}&=&
-\frac{\rmi}{2}\gamma^{\mu}(\partial_{\mu}\varphi^{x})\epsilon_{i}
-g\,P_{i}{}^{jx}\epsilon_{j},\label{gaugino}\\
\delta
\zeta^{A}&=&\frac{\rmi}{2}f_{X}^{iA}\gamma^{\mu}(\partial_{\mu}q^{X})
\epsilon_{i} - g\, \mathcal{N}^{iA}\epsilon_{i}.\label{hyperino}
\end{eqnarray}
Here, $\psi_{\mu}^{i}$, $\lambda_{i}^{x}$, $\zeta^{A}$ are the gravitini,
gaugini  and hyperini, respectively,
 $g$ denotes the gauge coupling, the $SU(2)$ connection $\upomega_{\mu}$ is
defined as $\omega_{\mu i}{}^{j}= (\partial_{\mu}q^{X}) \omega_{X
i}{}^{j} $, and
\begin{eqnarray}
P^{r}&=&h^{I}(\varphi)P_{I}^{r}(q), \label{Pdefinition}  \\
P^{rx}&=&-\sqrt{\frac{3}{2}}g^{xy}\partial_{y}P^r\\
\mathcal{N}^{iA}&=&\frac{\sqrt{6}}{4}f_X^{iA}(q)h^{I}(\varphi)K_{I}^{X}(q)\label{hshift}.
\end{eqnarray}
%
%
%
As usual, the potential is given by the sum of ``squares of the
fermionic shifts'' (the scalar expressions in the above
transformations of the fermions):
\begin{equation}
\mathcal{V}=-4P^{r}P^{r} +2P^{xr}P^{yr}g_{xy}+2\mathcal{N}^{iA}
\mathcal{N}^{jB}\varepsilon _{ij}C_{AB}, \label{scalarpotential}
\end{equation}
where $C_{AB}$ is the (antisymmetric) symplectic metric of $USp(2n_{H})$.
Using (\ref{KillingP}) and the quaternionic identity (\ref{quaterid}),
the scalar potential for vector and hypermultiplets can be written in the
form
\begin{equation}
\mathcal{V}\unity_{2} =4 \mathbf{P}^2 -3
(\partial_{x}\mathbf{P})(\partial^{x}\mathbf{P})
-(D_X\mathbf{P})(D^{X}\mathbf{P})\label{VPP2}.
\end{equation}
One clearly sees that the scalars of the vector- and hypermultiplets
enter the supersymmetry transformations and the scalar potential in
a rather different way.

\subsection{ Curved and flat BPS-domain walls }

In this paper, we are interested in Minkowski-sliced (``flat'')
 and AdS-sliced (``curved'')  domain walls
of the form
\begin{equation}
\rmd s^2=\rme^{2U(r)}g_{mn}(x)\,\rmd x^m \,\rmd x^n + \rmd r^2
\label{curvedmetric}
\end{equation}
with $g_{mn}(x)$ being either the 4D Minkowski-metric or
  a metric of  $AdS_{4}$ with curvature scale
$L_{4}$. In a curved domain wall background of the form
(\ref{curvedmetric}), when the scalar fields only depend on the
radial coordinate $r$, the vanishing of the supersymmetry
variations (\ref{gravitino})-(\ref{hyperino}) implies
\begin{eqnarray}
  \left[
\nabla_{m}^{AdS_4} +\gamma_{m} \left( \frac{1}{2} U^{\prime}\gamma_{5}
-\frac{ig}{\sqrt{6}}\mathbf{P} \right) \right] \epsilon &=&0,
\label{realgravitino1}\\
\left[D_{r}   + \gamma_{5} \left( -\frac{\rmi g}{\sqrt{6}} \mathbf{P}
\right)
\right] \epsilon &=&0,  \label{realgravitino2} \\
\left[  \gamma_{5}\varphi^{x\prime} + \rmi
g\,\sqrt{6}\,g^{xy}\partial_{y} \mathbf{P}
\right] \epsilon &=&0, \label{realgaugino}\\
 f_{X}^{iA}\left[\gamma_{5}q^{X\prime}- \rmi g \sqrt{\frac{8}{3}}R^{rXY}D_Y P^r \right]
\epsilon_{i}   &=& 0, \label{realhyperino}
\end{eqnarray}
where
\begin{equation}
D_r\epsilon_{i}\equiv \partial_{r}\epsilon_{i}
 -q^{X\prime}\omega_{Xi}{}^{j} \epsilon_{j} \label{Dr}
\end{equation}
has been introduced.
The gaugino variation suggests a spinor projector of the form
\footnote{As it turns out to be more convenient for the
$\mathcal{N}=4$ case, our $\mathbf{\Theta}$ differs by a factor $i$
from the one used in \cite{CCDVZ}: $\mathbf{\Theta}^{\textrm{here}} =
i \mathbf{\Theta}^{\textrm{there}}$.}
\begin{equation}
 \ifiTheta
  \epsilon _i=-\gamma _5\Theta _i{}^j\epsilon _j    \Leftrightarrow
\left(\unity_{2} +\gamma_{5}\mathbf{\Theta}\right)\epsilon =0,
  \else
  \epsilon _i=-\rmi\gamma _5\Theta _i{}^j\epsilon _j    \Leftrightarrow
\left(\unity +\rmi\gamma_{5}\mathbf{\Theta}\right)\epsilon =0,
  \fi
 \label{projepsilon}
\end{equation}
where $\mathbf{\Theta} ^2=\ifiTheta \else -\fi\unity_{2}   \Leftrightarrow
\Theta^r \Theta^r =\ifiTheta -\else \fi 1$.
Using this projector, the gaugino and hyperino BPS-conditions can be
brought to the following form \cite{CCDVZ}
\begin{eqnarray}
ig_{yx}\,\varphi^{x\prime}\mathbf{ \Theta} +
\sqrt{6}\,g\,\partial_{y}\mathbf{P}&=&0 ,
\label{vectorBPS1}\\
\ifiTheta\rmi\,\else\fi
g_{YX}q^{X\prime}\mathbf{\Theta}+\ifiTheta\rmi\else\fi
q^{X\prime}[\mathbf{R}_{YX},\mathbf{\Theta}] \ifiTheta +\else -\fi
\sqrt{6}\,g\, D_{Y}\mathbf{P} &=&0.
\label{BPS1}
\end{eqnarray}
The hyperino BPS-equation (\ref{BPS1}) can be written in the
equivalent form
\begin{equation}
\sqrt{6}\,g\,K_{Y}\ifiTheta+ 2\rmi\else-2\fi \, q^{X\prime} \{
\mathbf{R}_{YX},\mathbf{\Theta} \}=0  \label{BPS2}
\end{equation}
by contracting (\ref{BPS1}) with the $SU(2)$-curvature.

Contracting now (\ref{vectorBPS1}) and (\ref{BPS1}) with,
respectively,
 $\varphi^{y \prime}$
and $q^{Y\prime}$, one can solve for the projector $\mathbf{\Theta}$:
\cite{LopesCardoso:2001rt,Cardoso:2002ec,Behrndt:2002ee}
\begin{eqnarray}
\mathbf{\Theta}&=&\ifiTheta\rmi\else\fi g\,\sqrt{6}\,
\frac{\varphi^{x\prime}\partial_{x}\mathbf{P}}{\varphi^{y\prime}
\varphi^{z\prime}g_{yz}}\label{Thetavector}\\
\mathbf{\Theta}&=&\ifiTheta\rmi\else\fi g\,\sqrt{6}\,\frac{q^{X\prime}D_{X}
\mathbf{P}}{q^{Y\prime} q^{Z\prime} g_{YZ}}\label{Thetahyper}.
\end{eqnarray}
When both vector multiplet scalars and hyper-scalars are non-trivial,
consistency of (\ref{Thetahyper}) and (\ref{Thetavector}) obviously
requires
\begin{equation}\label{consistency}
\frac{q^{X\prime}D_{X}\mathbf{P}}{q^{Y\prime}q^{Z\prime}g_{YZ}}=
\frac{\varphi^{x\prime}\partial_{x}\mathbf{P}}{\varphi^{y\prime}
\varphi^{z\prime}g_{yz}}.
\end{equation}
Squaring (\ref{Thetavector}) and   (\ref{Thetahyper})
  finally yields the equations of motion
for the scalar fields,
\begin{eqnarray}
\varphi^{x\prime}\varphi^{y\prime}g_{xy}&=&\pm g\sqrt{6}
\sqrt{-(\varphi^{x\prime}\partial_{x}\mathbf{P})^2}\\
q^{X\prime}q^{Y\prime}g_{XY}&=&\pm g\sqrt{6}
\sqrt{-(q^{X\prime} D_{X}\mathbf{P})^2}.
\end{eqnarray}
As for the warp factor $U(r)$,
a first order equation can be
obtained from the integrability condition of (\ref{realgravitino1}),
which yields
\begin{equation}
(U^{\prime})^{2}= -\frac{e^{-2U}}{L_{4}^2} -\frac{2}{3}g^2\mathbf{P}^2.\label{Hans}
\end{equation}
However, the compatibility condition of (\ref{realgravitino1})
and (\ref{projepsilon}) also implies a first order equation for $U(r)$:
\begin{equation}
U^{\prime}=-\frac{ig}{\sqrt{6}} \{          \mathbf{\Theta}, \mathbf{P}   \}. \label{Franz}
\end{equation}
Consistency of (\ref{Hans}) and (\ref{Franz}) then implies an algebraic
equation for the warp factor:
\begin{equation}
\frac{6 e^{-2U}}{g^2 L_{4}^{2}}\unity_{2}= \{ \mathbf{\Theta}, \mathbf{P}
\}^{2} -4 \mathbf{P}^2  \label{algebraic}.
\end{equation}
This is an important equation, because it tells us that the domain
wall is flat (corresponding to $L_{4}    \rightarrow \infty$) if and
only if $\mathbf{P}$ and $\mathbf{\Theta}$ are proportional to one
another, $\mathbf{P}=c \mathbf{\Theta}$.

There is yet
one other important consistency condition,
which follows from the compatibility of (\ref{realgravitino2})
and (\ref{projepsilon}).  It reads
\begin{equation}
  \left[\mathbf{\Theta }\,,\,D_r \mathbf{\Theta}  +\ifiTheta\rmi\else\fi\sqrt{\frac23}g\,
\mathbf{P}\right]=0.
 \label{commutator1}
\end{equation}
Since eqs.  (\ref{Thetavector}) and (\ref{Thetahyper})  imply
that $\mathbf{\Theta}$ is proportional to $D_r\mathbf{P}$:
\begin{equation}
  D_r\mathbf{P}\equiv \varphi ^{\prime x}\partial _x\mathbf{P}+q^{\prime
  X}D_X\mathbf{P}= \ifiTheta-\frac \rmi{\sqrt 6 g}\else \frac 1{\sqrt 6 g}\fi
   g_{\Lambda\Sigma}\phi^{\Lambda\prime}\phi^{\Sigma\prime}
  \mathbf{\Theta},
 \label{DrPproptoTheta}
\end{equation}
where $\phi^{\Lambda}=\{\varphi^{x}, q^{X}\}$,  the
 consistency condition (\ref{commutator1}) can be rewritten in the
form
\begin{equation}
  \left[ D_r\mathbf{P},D_rD_r\mathbf{P}
  +\frac 13 g_{\Lambda\Sigma}\phi^{\Lambda\prime}\phi^{\Sigma\prime}\mathbf{P}\right]
  =0.
 \label{commutatorD}
\end{equation}

Obviously, (\ref{commutatorD})
  is a constraint on the possible field dependence of
$\mathbf{P}$ on  a supersymmetric domain wall solution. As was shown
in \cite{CCDVZ}, this constraint is only partially compatible with
the geometric constraints from very special geometry. More
precisely, if the domain wall is supported only by vector multiplet
scalars, (\ref{commutatorD}) can only be satisfied if
$\mathbf{\Theta}$ and $\mathbf{P}$ are proportional to one another.
But, according to (\ref{algebraic}), this means that     the domain
wall then   has to be \emph{flat}. Thus, any BPS-domain wall that is
supported by vector multiplet
 scalars only has to be flat, and curved domain walls require non-trivial hyperscalar profiles \cite{CCDVZ}.




\subsection{The relation to ($\mathcal{N}=2$)  ``fake'' supergravity}

The BPS-domain wall solutions reviewed in the previous subsection
are classically stable solutions of the underlying gauged
supergravity theories. This follows from standard arguments based on
the existence of Killing spinors and the first order form of the
field equations   along the lines   of \cite{Witten:1981mf,JN}. In
\cite{Townsend:1984iu,Skenderis:1999mm,DeWolfe:1999cp}, 
these stability arguments were formalized and generalized to flat
domain wall solutions of a broader class of theories which, while
having some superficial  similarities with true supergravity
theories, do not necessarily have to be
  supersymmetric and can live in any space-time dimension $D=(d+1)$.
In ref. \cite{Freedman:2003ax}, such theories were dubbed ``fake''
supergravity theories, and the formalism was further generalized and refined
to
include also curved domain walls. More precisely, the theories
studied in ref. \cite{Freedman:2003ax} are  gravitational theories
with a single scalar field $\phi$ and an action
\begin{equation}
S=\int d^{d+1}x \sqrt{-g} \left[ \frac{1}{2\kappa^{2}}R-\frac{1}{2}
\partial_{\mu}\phi\partial^{\mu}\phi -V(\phi) \right], \label{Freedman-1}
\end{equation}
with a scalar potential $V(\phi)$ given by
\begin{equation}
V(\phi)=\frac{2(d-1)^2}{\kappa^2}\left(\frac12\trace\right) \left[
\frac{1}{\kappa^2}(\partial _\phi \mathbf{W})^2
-\frac{d}{d-1}\mathbf{W}^{2} \right]. \label{Freedman0a}
\end{equation}
Here, $\mathbf{W}(\phi)$ is an $\mathfrak{su}(2)$-valued $(2\times
2)$-matrix, which implies that quadratic expressions such as
$\mathbf{W}^2$, $(\partial_{\phi}\mathbf{W})^2$ or
$\{\mathbf{W},\partial _\phi \mathbf{W}\}$ are proportional to the
unit matrix. This allows one to write the potential in an equivalent
form without explicitly taking the trace:
\begin{equation}
V(\phi)\unity_{2}=\frac{2(d-1)^2}{\kappa^2} \left[
\frac{1}{\kappa^2}(\partial _\phi \mathbf{W})^2
-\frac{d}{d-1}\mathbf{W}^{2}     \right].   \label{Freedman0}
\end{equation}

The matrix $\mathbf{W}$ also enters some ``fake'' Killing spinor
equations for an $SU(2)$-doublet spinor  $\epsilon$,
\begin{eqnarray}
\left[ \nabla_{m}^{AdS_{d}} +\gamma_{m} \left( \frac{1}{2}
\,U^{\prime}\gamma_{5}
+\mathbf{W} \right) \right] \epsilon &=& 0, \label{Freedman1}\\[2mm]
\left[ \partial_{r}+\gamma_{5}\mathbf{W}\right]
\epsilon &=& 0, \label{Freedman2}\\[2mm]
 \left[ \gamma_{5} \phi^{\prime} - \frac{2(d-1)}{\kappa^2}\partial _\phi
 \mathbf{W}\right] \epsilon &=& 0. \label{Freedman3}
\end{eqnarray}
In this expression,
$U(r)$ is the warp factor of   a $(d+1)$-dimensional     metric of the form 
(\ref{curvedmetric}), and
$\nabla_{m}^{AdS_d}$ denotes the
covariant derivative for the $AdS_d$ background metric $g_{mn}(x)$.
The
prime means 
a derivative with respect to $r$, which we have chosen, for all $d$, to be the fifth coordinate $x^5$.
These fake Killing spinor equations can be thought of as arising from some
``fake'' supersymmetry
transformation rules  in a domain wall background (\ref{curvedmetric}),
\begin{eqnarray} \left[ \nabla_{\mu}
+\gamma_{\mu} \mathbf{W} \right] \epsilon &=& 0, \nonumber\\
[2mm]
 \left[ \gamma^\mu \nabla_\mu \phi - \frac{2(d-1)}{\kappa^2}\partial _\phi
 \mathbf{W}\right] \epsilon &=& 0, \label{Freedman000}
\end{eqnarray}
where $\nabla_\mu\epsilon = \left(\partial_\mu + \frac14
\,{\omega_\mu}^{\nu\rho} \gamma_{\nu\rho}\right)\epsilon$.

It is shown in \cite{Freedman:2003ax} that the system
(\ref{Freedman1})-(\ref{Freedman3}) reproduces the second order field
equations for the warp factor $U(r)$ and the scalar field $\phi(r)$ that
follow from (\ref{Freedman-1}) and (\ref{Freedman0a}) with
\begin{equation}
\frac{e^{-2U(r)}}{L_{d}^{2}}=\frac{2\textrm{Tr}\mathbf{W}^{2}\textrm{Tr}(\partial_{\phi}\mathbf{W})^2-
\textrm{Tr} \{ \mathbf{W},\partial_{\phi}\mathbf{W}
\}^{2}}{\textrm{Tr}(\partial_{\phi}\mathbf{W})^{2}}
\end{equation}
(where $L_d^2=-12/R_{AdS}$ is determined by the scalar curvature of the
AdS space) \emph{provided that} the ``superpotential'' $\mathbf{W}(\phi)$
satisfies the constraint
\begin{equation}
\left[\partial _\phi  \mathbf{W}, \frac{d-1}{\kappa^2}
\partial _\phi\partial _\phi\mathbf{W} +\mathbf{W} \right]=0, \label{Freedman16}
\end{equation}
which is a compatibility condition of (\ref{Freedman2}) and
(\ref{Freedman3}).

As there are some obvious similarities with the analogous equations in
sections 2.1 and 2.2, one might wonder what exactly the relation
between fake and real supergravity is, and how far-reaching the
similarities are. As was found in \cite{CCDVZ}, the answer to this
question turns out to be  surprisingly simple.
In order to see this, three cases should  be distinguished:
\begin{enumerate}
\item
 The domain wall is supported only by scalar fields $\varphi^x$ that sit in vector multiplets.
\item  The domain wall is supported only  by scalar fields $q^X$
  that sit in hypermultiplets.
\item The domain wall is supported by both types of scalar fields, $\varphi^{x}$ and $q^X$.
\end{enumerate}

Let us first consider case (i). In this case,      a supersymmetric
domain wall solution is  given by profile functions $U(r)$ and
$\varphi^{x}(r)$ that solve the BPS-equations
(\ref{realgravitino1})-(\ref{realgaugino}), where now $D_r
\epsilon_i = \partial_{r} \epsilon_{i}$, because $q^{X \prime} =0$:
\begin{eqnarray}
  \left[
\nabla_{m}^{AdS_4} +\gamma_{m} \left( \frac{1}{2} U^{\prime}\gamma_{5}
-\frac{ig}{\sqrt{6}}\mathbf{P} \right) \right] \epsilon &=&0,
\label{realgravitino1a}\\
\left[\partial_{r}   + \gamma_{5} \left( -\frac{\rmi g}{\sqrt{6}} \mathbf{P}
\right)
\right] \epsilon &=&0,  \label{realgravitino2a} \\
\left[  \gamma_{5}\varphi^{x\prime} + \rmi
g\,\sqrt{6}\,g^{xy}\partial_{y} \mathbf{P}
\right] \epsilon &=&0. \label{realgauginoa}
\end{eqnarray}
Obviously, the two gravitino equations (\ref{realgravitino1a}) and (\ref{realgravitino2a}) are now exactly of the fake supergravity form
(\ref{Freedman1}) and  (\ref{Freedman2}) if we identify
\begin{equation}
\mathbf{W} = -\frac{\rmi g}{\sqrt{6}} \mathbf{P}.
 \label{WP}
\end{equation}
Upon this identification, the gaugino equation (\ref{realgauginoa})
also assumes the form (\ref{Freedman3}),
the only difference being
the different number of scalar fields in these two expressions.
There are now two attitudes one could take. One
  could, for example, simply      view (\ref{realgauginoa}) as a suggestion for
a generalized form of fake supergravity which  involves several
scalar fields. As we will see, however, running hypermultiplet
scalars in cases (ii) and (iii)   suggest  quite  a different
generalization to several scalar fields. We will therefore, at this
point, choose the interpretation adopted in \cite{CCDVZ} and bring
(\ref{realgauginoa}) and (\ref{Freedman3}) to exact agreement, by
reducing (\ref{realgauginoa}) effectively to an equation for one
scalar field. In order to do this, one recalls that a given domain
wall solution defines a curve on the scalar manifold $\mathcal{M}$,
which in the case at hand lies entirely in $\mathcal{M}_{VS}$. As
the coordinates $\varphi^{x}$ on $\mathcal{M}_{VS}$ can be chosen at
will, one can, at least locally, choose ``adapted'' coordinates
$\varphi^{x}(r)=(\varphi(r),\varphi^{\hat{x}})$, where $\varphi(r)$
is aligned with the flow curve, and the other scalars
$\varphi^{\hat{x}}$ correspondingly do not depend on $r$. It is
convenient (and locally always possible)  to choose these
$r$-independent   coordinates $\varphi^{\hat{x}}$ to be orthogonal
to the coordinate $\varphi$, at least on the flow curve $\varphi(r)$
itself (or on a sufficiently short segment of it). This is illustrated in Figure 1.
\begin{figure}
\begin{picture}(100,152)
\put(120,10){$\mathcal{M}_{\rm VS}$}
\put(325,10){$\mathcal{M}_{\rm VS}$}
\put(45,125){$\varphi^{1}$}
\put(105,75){$\varphi^{2}$}
\put(280,35){$\varphi^{\hat{x}}$}
\put(312,112){$\varphi$}
\put(15,-15){a) Generic coordinates}
\put(215,-15){b) Adapted coordinates}
\resizebox{12cm}{5cm}{\includegraphics{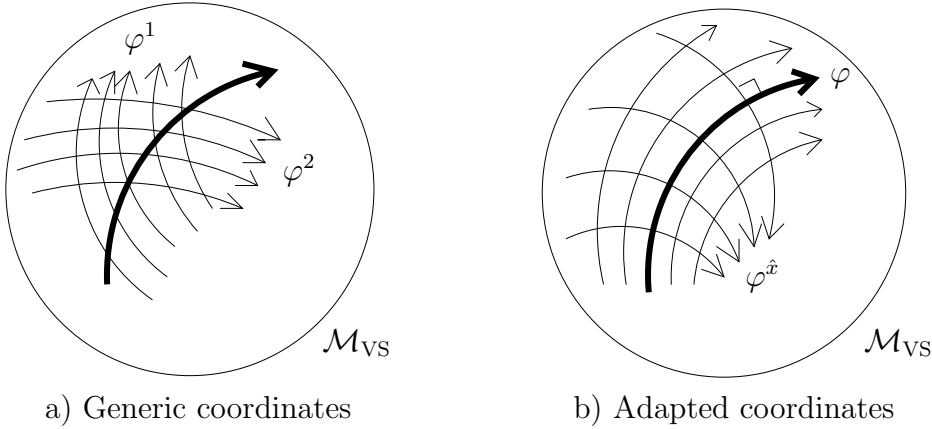}}
\end{picture}
\begin{center}
\caption[Figure 1:]{
A given               domain wall   defines a  flow   curve  (thick arrow) 
on the scalar manifold $\mathcal{M}_{\rm VS}$. 
In  a), the thin  arrows  correspond to a generic coordinate system $\varphi^{x}= (\varphi^1,\varphi^2)$.  In     b),    the coordinate system $\varphi^{x}=(\varphi,\varphi^{\hat{x}})$  is 
adapted to the flow curve, i.e., the  flow curve coincides with a coordinate line of $\varphi$ and intersects the coordinate lines $\varphi^{\hat{x}}$ at 
right angles.   }
\end{center}
\end{figure}
On the flow
curve, the scalar field metric $g_{xy}$ then takes the form
\begin{equation}
g_{xy}= \left( \begin{array}{cc}
g_{\varphi\varphi} & 0 \\
0 & g_{\hat{x}\hat{y}}
\end{array} \right). \label{adaptedmetric}
\end{equation}
By a suitable rescaling of $\varphi$, one  can, on the curve
$(\varphi(r),\varphi^{\hat{x}})$, also achieve $g_{\varphi \varphi}
=1$.
The $\varphi$-component of the  gaugino equation
(\ref{realgauginoa})  now coincides with the fake supergravity
version  (\ref{Freedman3}), and the orthogonal components of
(\ref{realgauginoa}) imply
\begin{equation}
\partial_{\hat{x}}\mathbf{P}=0.  \label{hatP}
\end{equation}

As $q^{X\prime}=0$ also implies $D_{X}\mathbf{P}=0$ via
(\ref{BPS1}), the effective
  scalar potential (\ref{VPP2}) on the domain wall assumes a simple form,
\begin{eqnarray}
V\unity_{2} =  g^2 \mathcal{V}\unity_{2}  &=& 4g^2 \mathbf{P}^2 -3g^2
(\partial_{\varphi}\mathbf{P})^{2}\nonumber\\
&=& -24 \mathbf{W}^2 +18(\partial_{\varphi}\mathbf{W})^{2} \label{VPPsimple},
\end{eqnarray}
which precisely matches (\ref{Freedman0}) for $d=4$, $\kappa=1$ and
$\phi=\varphi$.
Thus, once the $(n_{V}-1)$ orthogonal BPS-equations (\ref{hatP})
have determined the line of flow on the scalar manifold, the
effective dynamics of the supporting scalar field and the warp
factor are precisely described by  single-field ``fake''
supergravity equations  \`{a} la \cite{Freedman:2003ax}.

Let us now turn to case (ii) and assume the domain wall is supported
by hypermultiplet scalars only. In that case, the gaugino BPS-equation
(\ref{realgaugino}) implies
\begin{equation}
\partial_{x}\mathbf{P}=0\label{dxP=0},
\end{equation}
because we now have $\varphi^{x \prime}=0$.
The gravitino equations (\ref{realgravitino1}) and (\ref{realgravitino2})
are again  of the same form as the corresponding fake supergravity equations
(\ref{Freedman1}) and (\ref{Freedman2}) provided that we again make
the identification (\ref{WP}) and gauge away the $SU(2)$ connection
along the flow line:
\begin{equation}
q^{X \prime}\omega_{Xi}{}^{j}=0 \qquad   (SU(2) \textrm{ gauge choice})   , \label{SU2gauge}
\end{equation}
which is locally always possible, as explained in \cite{CCDVZ}.
However, due to the explicit
  appearance of the $SU(2)$ curvature tensor,
the hyperino  BPS-condition (\ref{realhyperino}), or its equivalent version
 (\ref{BPS1}),
\begin{equation}
\ifiTheta\rmi\,\else\fi
g_{YX}q^{X\prime}\mathbf{\Theta}+\ifiTheta\rmi\else\fi
q^{X\prime}[\mathbf{R}_{YX},\mathbf{\Theta}] \ifiTheta +\else -\fi
\sqrt{6}\,g\, D_{Y}\mathbf{P} =0,
\label{BPS1a}
\end{equation}
clearly differs
from the corresponding fake supergravity analogue
(\ref{Freedman3}). Likewise, the scalar potential  (\ref{VPP2})
now reads (remembering (\ref{dxP=0}))
\begin{equation}
\mathcal{V}\unity_{2} =4 \mathbf{P}^2
-(D_X\mathbf{P})(D^{X}\mathbf{P})\label{VPP2a},
\end{equation}
which seems to have the ``wrong'' prefactor in front of the derivative term
when  compared to  (\ref{Freedman0}).
In order to make contact between the two formulations, one again has
to interpret fake supergravity as the effective theory along the
flow line. More precisely, for a given domain wall solution, one
again   chooses adapted coordinates $q^{X}(r)=(q(r),q^{\hat{X}})$
such that, on the flow curve,
\begin{equation}
g_{XY}= \left( \begin{array}{cc}
g_{qq} & 0 \\
0 & g_{\hat{X}\hat{Y}}
\end{array} \right), \label{adaptedmetric2}
\end{equation}
where $g_{qq}(q(r),q^{\hat{X}})$ can again be chosen to be equal to one. The
supersymmetry condition (\ref{BPS1a}) now splits into two equations:
\begin{eqnarray}
q^{\prime}\mathbf{\Theta} -\ifiTheta\rmi\,\else\fi g\,\sqrt{6}D_{q}\mathbf{P}&=&0,\label{hyper1}\\
q^{\prime}[\mathbf{R}_{\hat{X}q},\mathbf{\Theta}]-\ifiTheta\rmi\,\else\fi
g\,\sqrt{6} D_{\hat{X}}\mathbf{P}&=&0.\label{hyper2}
\end{eqnarray}
In view of (\ref{projepsilon}), the first equation (\ref{hyper1}) is
easily seen to be equivalent to the fake supergravity equation
(\ref{Freedman3}) provided the $SU(2)$ gauge (\ref{SU2gauge}) is
imposed. The second equation should again  be viewed as a set of
constraint equations that determines the position of the flow curve
in the full scalar manifold  $\mathcal{M}_{Q}$ as a co-dimension one
hypersurface. Note that (\ref{hyper2}) is different from the
analogous equation  (\ref{hatP})   in the case of running vector
multiplet scalars, as it no longer implies that the hatted
derivatives of $\mathbf{P}$ have to vanish. In fact, one can show
that (\ref{hyper2})        implies that, on a BPS-domain wall
solution \cite{CCDVZ},
\begin{equation}
D_{\hat{X}}\mathbf{P} D^{\hat{X}}\mathbf{P}=  2 D_{q} \mathbf{P} D^{q}
\mathbf{P}, \label{Phatq}
\end{equation}
showing that at least some components of $D_{\hat{X}}\mathbf{P}$ have to be non-zero. Luckily, this is precisely as it should be, because (\ref{Phatq}) exactly corrects the ``wrong'' prefactor $(-1)$ in the potential (\ref{VPP2a})
to $(-3)$, so that, using the $SU(2)$ gauge (\ref{SU2gauge}),
\begin{eqnarray}
V\unity_{2} =  g^2 \mathcal{V}\unity_{2}  &=& 4g^2 \mathbf{P}^2 -3g^2
(\partial_{q}\mathbf{P})^{2}\nonumber\\
&=& -24 \mathbf{W}^2 +18(\partial_{q}\mathbf{W})^{2} \label{VPPsimple2},
\end{eqnarray}
i.e., exactly as in ``fake'' supergravity.

Let us finally mention the last case (iii) of running  vector- and
hypermultiplet scalars. The flow curve now has non-trivial projections
$\varphi^{x}(r)$ and $q^{X}(r)$
on both scalar manifold components, $\mathcal{M}_{VS}$ and $\mathcal{M}_{Q}$.
One can    then, in a first step, choose separate adapted coordinates
  $\varphi^{x}(r)=(\varphi(r),\varphi^{\hat{x}})$ and $q^{X}(r)=(q(r),q^{\hat{X}})$
   on  $\mathcal{M}_{VS}$ and $\mathcal{M}_{Q}$, respectively.
In the       $SU(2)$-gauge (\ref{SU2gauge}),
the BPS-equations and the scalar potential then look the same for both types of
 scalars $\varphi$ and $q$. One can then employ a coordinate transformation
in the      $(\varphi,q)$-plane,
\begin{equation}
(\varphi(r),q(r)) \rightarrow (\phi(r),\hat{\phi}),
 \label{coordtrphi}
\end{equation}
 such that, locally,
$\partial_{r}=q^{\prime}\partial_{q}+\varphi^{\prime}\partial_{\varphi}=
\phi^{\prime}\partial_{\phi}$. In this new, totally adapted
coordinate system, the BPS-equation for $\phi$ and the scalar
potential as a function of $\phi(r)$ then take the standard ``fake''
supergravity form \cite{CCDVZ}.

The lesson we learn from this is that a generalization of the
single-field formalism of fake supergravity to several scalar fields
is not so straightforward, as the prefactors in the scalar potential
can be different and non-trivial connections and curvatures might
come into play. However, interpreting   single-field          fake
supergravity as an effective theory along the flow curve seems to
make sense in all cases. It is this latter interpretation that we
will now try to  generalize to  the $\mathcal{N}=4$ case.




\section{BPS-domain walls in $\mathcal{N}=4$ fake and true supergravity}

In this section, we study   curved and flat BPS-domain walls      in
5D, $\mathcal{N}=4$ gauged supergravity  and verify to what extend
one can generalize ``$\mathcal{N}=2$'' fake supergravity to
``$\mathcal{N}=4$'' fake supergravity. We begin with a brief summary
of       5D, $\mathcal{N}=4$ ungauged \cite{Awada} and gauged
\cite{Romans,DHZ} supergravity.  Our notation follows that of ref.
\cite{DHZ}, to which the reader is referred for further details.

\subsection{ Ungauged 5D, $\mathcal{N}=4$ supergravity}
In the previous section, the index $i=1,2$ was used  to denote the
fundamental representation of the R-symmetry group $Usp(2) \cong SU(2)$
of the 5D, $\mathcal{N}=2$ Poincar\'{e} superalgebra. In this section,
\begin{equation}
i=1,\ldots,4
\end{equation}
will instead
denote the fundamental representation of the $\mathcal{N}=4$ R-symmetry
group $Usp(4)$, which    is locally isomorphic to $SO(5)$.

In \emph{ungauged} 5D supergravity, vector fields and antisymmetric
tensor fields are equivalent, and the most general ungauged
$\mathcal{N}=4$ theory describes the coupling of $n$ vector
multiplets to supergravity.
%

%
%
%
%
%
%
%
%
%
 The supergravity
 multiplet,
 \begin{equation}
\Big(\,\,e_\mu{}^m\,,~\psi_\mu^i\,,~A_\mu^{ij}\,,~a_\mu\,,
~\chi^i\,,~\sigma\,\Big), \label{n=4_sugra}
\end{equation}
contains the graviton $e_\mu{}^m$, four gravitini $\psi_\mu^i$,
six vector fields $(A_\mu^{ij},a_\mu)$, four spin $1/2$ fermions
$\chi^i$ and one
 real scalar field $\sigma$.
The vector field
$a_\mu$ is $USp(4)$ inert, whereas the vector fields $A_\mu^{ij}$
transform in the {\bf 5} of $USp(4)$, i.e.,
\begin{equation}
A_\mu^{ij} \ = \ -A_\mu^{ji}~,\qquad A_\mu^{ij}\,\Omega_{ij} \ = \
0, \label{symplectictraceless}
\end{equation}
with $\Omega_{ij}$ being the symplectic metric of $USp(4)$.

An $\cN=4$ vector multiplet is given by
\begin{equation}
\Big(\,A_\mu\,,~\lambda^i\,,~\varphi^{ij}\,\Big),
\end{equation}
where $A_\mu$ is a vector field, $\lambda^i$ denotes four spin $1/2$
fields, and the $\varphi^{ij}$ are     scalar fields transforming in
the {\bf 5} of $USp(4)$, similar to eq. (\ref{symplectictraceless}).
Coupling $n$ vector multiplets to supergravity, the field content
of the theory can then  be summarized as follows
\begin{equation}
\Big(\,e_\mu{}^m\,,~\psi_\mu^i\, ,~A_\mu^\tI \, ,~a_\mu\,,~\chi^i\,,~\la^{ia}\,,~\si\,,~\varphi^x\,\Big). \label{fullcontent}
\end{equation}
Here, $a=1,\ldots,n$ counts the number of vector multiplets
whereas
 $\tI=1,\ldots,(5+n)$  collectively denotes
the $A_\mu^{ij}$ and the vector fields of the vector multiplets.
Similarly, $x=1,\ldots,5n$ is a collective index for all  the scalar
fields in the vector multiplets. We will further adopt the following
convention to raise and lower $USp(4)$ indices:
\begin{equation}
T^i \ = \ \Omega^{ij}\,T_j~,\quad T_i \ = \ T^j\,\Omega_{ji},
\label{summationconvention}
\end{equation}
whereas $a,b$ are raised and lowered with $\delta^{ab}$. Before we
proceed, we note that, in  a more familiar language, quantities such
as $A_{\mu i}{}^{j}$  in the $\mathbf{5}$ of $Usp(4)\cong SO(5)$ can
be expressed as $A_{\mu i}{}^{j}= A_{\mu}^{\alpha}\Gamma_{\alpha
i}{}^{j}$, where $\alpha=1,\ldots, 5$, and $\Gamma_{\alpha i}{}^{j}$
denote $SO(5)$ gamma matrices,
\begin{equation}
\Gamma_{\alpha i}{}^{j}\Gamma_{\beta j}{}^{k} +(\alpha
\leftrightarrow \beta) = 2\delta_{\alpha \beta } \delta_{i}^{k}
\label{SO5Gamma}.
\end{equation}
As was shown in \cite{Awada}, the manifold spanned by the $(5n+1)$
scalar fields is
\begin{equation}
    \label{cM}
\cM \ = \ \frac{SO(5,n)}{SO(5)\times SO(n)}\times SO(1,1),
\end{equation}
where the $SO(1,1)$ part corresponds to the $USp(4)$-singlet
$\sigma$ of the supergravity multiplet. The theory  therefore has  a
{\em global} symmetry group $SO(5,n)\times SO(1,1)$ and a {\em local
composite} $SO(5)\times SO(n)$ invariance.
The coset part of the scalar manifold $\cM$ can be described in two
different  ways:
\begin{enumerate}
\item  \emph{Standard sigma model description:}\\
As in (\ref{fullcontent}) one can simply
  choose a
parameterization in terms of $5n$ independent
real   coordinates $\varphi^x$ on the coset space.
The vielbeins on the coset space can then be chosen to be of the form
\begin{equation}
f_{x}^{ija} =-f_{x}^{jia}, \qquad f_{x}^{ija}\Omega_{ij}=0,
\end{equation}
where $[ij]$ and $a$ refer to the natural $\textbf{(5,n)}$  tangent
space decomposition with respect to the   holonomy group
$\cH=SO(5)\times SO(n)$. The inverse vielbein, $f_{aij}^{x}$, is
defined by
\begin{equation}
f_{x}^{aij} f_{kl}^{xb}=4
\Big(\delta_{k}^{[i}\delta_{l}^{j]}-\frac{1}{4}\Omega^{ij}\Omega_{kl}
\Big) \delta^{ab}. \label{finverse}
\end{equation}
 The non-linear  $\sigma$-model  metric
$g_{xy}$ on the coset part of $\cM$ is then   given   by
\begin{equation}
g_{xy} \ = \ \frac14\,f_x^{ija}\,f^a_{yij},
\end{equation}
and the kinetic term for the scalar fields takes the standard form
 $\frac12 g_{xy}\prt_\mu \varphi^{x}\prt^\mu \varphi^y$.
  This way of describing $\cM$ is particularly useful
 for discussing geometrical properties of the theory.\\
\item \emph{Coset representatives:}\\
The parametrization that makes the symmetries of the theory as
manifest as possible is in terms of coset representatives, i.e.,
$(5+n)\times (5+n)$ matrices $\,L_\tI{}^A\,  \subset \cG \equiv
SO(5,n)$ that are subject to local (``composite'') $\cH=SO(5)\times
SO(n)\,$ transformations (acting on the index $A$)  and admit the
action of global $\cG=SO(5,n)\,$ transformations (acting   on  the
index $\tilde{I}$). The index $A=1,\ldots,(5+n)$ naturally
decomposes into $A=(ij,a)$, and so do the coset representatives,
$L_{\tI}{}^{A}=(L_{\tI}^{ij},L_{\tI}^{a})$, where $L_{\tI}^{ij}$
transforms in the $\mathbf{5}$ of $SO(5)$, just as in
(\ref{symplectictraceless}).
 Denoting the
inverse of $\,L_\tI{}^A\,$ by $\,L_A{}^\tI\,$ (i.e., $\,L_\tI{}^A\,
\,L_B{}^\tI\ = \delta_{B}^{A}$), the vielbeins on $\cG/\cH$ and the
composite $\cH$-connections are determined from the $\cG$-invariant
1-form:
\begin{equation}
L^{-1}dL \ = \ Q^{ab}\,\mathfrak{T}_{ab} + Q^{ij}\,
\mathfrak{T}_{ij} + P^{aij} \, \mathfrak{T}_{aij},
\end{equation}
where $\,(\mathfrak{T}_{ab},\mathfrak{T}_{ij})\,$ are the
generators of the Lie algebra $\mathfrak{h}$ of $\cH$, and
$\mathfrak{T}_{aij}$ denotes the  generators of the coset part of
the Lie algebra $\mathfrak{g}$ of $\cG$. More precisely,
\begin{equation}\label{Q}
Q^{ab} \ = \ L^{\tI a}dL_\tI{}^b\qquad\mathrm{and}\qquad Q^{ij} \
= \ L^{\tI ik} dL_{\tI k}{}^j
\end{equation}
are the composite $SO(n)$ and $USp(4)$ connections, respectively,
and
\begin{equation}\label{P}
P^{aij} \ = \  L^{\tI a}
dL_{\tI}{}^{ij}=-\frac{1}{2}f_{x}^{aij}\,d\varphi^x
\end{equation}
describes the space-time pull-back of the $\cG/\cH$ vielbein. Note
that $Q_\mu^{ab}$ is antisymmetric in the $SO(n)$ indices, whereas
$Q_\mu^{ij}$ is symmetric in $i$ and $j$. Denoting by $D_x$ 
 the corresponding  $USp(4)$ and $SO(n)$ covariant derivative, 
one has the following differential realations  for the coset representatives
\cite{Awada}:
\begin{eqnarray}
    D_x L_{\tI ij} & = & -\frac{1}{2} L_{\tI}^a f_{xij}^a,\label{diff1}  \\
    D_x L^{\tI}_{ij} & = & \frac{1}{2} L^{\tI\,a} f_{xij}^a,\label{diff2}  \\
    D_x L^a_{\tI} & = & -\frac{1}{2} f_{xij}^a L_{\tI}^{ij}, \label{diff3} \\
    D_x L^{\tI a} & = & \frac{1}{2} f_{x}^{ij a} L^{\tI}_{ij},\label{diff4}.
\end{eqnarray}
We finally note the identities  (see       \cite{Awada,DHZ})              
\begin{equation}
\delta_{\tI}{}^{\tJ} \ = \ L_\tI^{ij}L_{ ij}^{\tJ}+L_\tI^aL^{\tJ a}~,\quad C_{\tI\tJ}
\ = \
L_\tI^{ij}L_{\tJ ij}-L_\tI^aL_\tJ^a, \label{dC}
\end{equation}
where $C_{\tI\tJ}$ is the (constant) $SO(5,n)$ metric.
\end{enumerate}
In the following, we will frequently switch between these two
formulations, which is easily done using eq.\ (\ref{P}).
%
%
%
The Lagrangians and supersymmetry transformation rules can be found
in \cite{Awada,DHZ}. 



\subsection{   5D, $\mathcal{N}=4$ gauged supergravity}
The above ungauged supergravity theories cannot support domain
walls, because their scalar potentials vanish identically, as
enforced by supersymmetry. As is typical   for extended
supergravity,  non-trivial scalar potentials are related to
non-trivial local gauge groups, $K$. These gauge groups cannot be
chosen at will, but have to be subgroups of the global symmetry
group $G=SO(1,1)\times SO(5,n)$ of  the     ungauged supergravity.
As is explained in more detail in \cite{DHZ}, the $SO(1,1)$ factor
in $G$ cannot be gauged, and all gauge groups, $K$, actually  have
to be suitable subgroups of $\mathcal{G}=SO(5,n)$. Under
$\mathcal{G}=SO(5,n)$, the vector fields $A_{\mu}^{\tI}$ transform
in the defining representation $\mathbf{(5+n)}$, whereas $a_{\mu}$
is $SO(5,n)$-inert. If some of these vector fields are promoted to
gauge fields of a local gauge group $K\subset SO(5,n)$ under which
some of the other fields are charged, the general equivalence
between vector and tensor fields is broken \cite{DHZ}.  Instead, one now has to
distinguish carefully between vector and tensor fields and pay
attention to their transformation properties
   under the gauge group $K\subset SO(5,n)$.
The result of the analysis in  ref. \cite{DHZ} is as follows
\footnote{In \cite{DHZ}, particular attention was paid to gauge
groups of the form $K=\textrm{ Abelian } \times \textrm{ semi-simple
}$, but all results of \cite{DHZ} equally apply to all gauge groups
$K\subset SO(5,n)$ under which the $\mathbf{(5+n)}$  of $SO(5,n)$
decomposes into a completely reducible representation so that tensor
fields and vector fields are not connected by $K$-transformations
   (see  also \cite{Bergshoeff:2004kh}). 
 We only consider such gauge groups in this paper.           They include, in particular, the gauge groups       encountered 
 in \cite{LouisMicu} and \cite{Zwirner}.                                             }:\\
\begin{enumerate}
\item     If the gauge group $K$ is a direct product of an Abelian
factor $K_{A}$ and a (possibly trivial)  non-Abelian factor $K_{S}$,
the
 Abelian factor $K_A$
 has to be
 one-dimensional (i.e., either $U(1)$ or $SO(1,1)$, but
no higher powers/products thereof). The gauge field of this Abelian
factor is $a_{\mu}$. Decomposing the vector fields $A_{\mu}^{\tI}$
into $K_{A}$ singlets, $A_{\mu}^{I}$, and non-singlets,
$A_{\mu}^{M}$, the non-singlets $A_{\mu}^{M}$ have to be converted
to tensor fields $B_{\mu\nu}^{M}$ for the gauging to be possible:
\begin{equation}
A_{\mu}^{\tI} \rightarrow (A_{\mu}^{I},B_{\mu\nu}^{M})
\end{equation}
\item   A possible non-Abelian factor, $K_S$, is gauged by the remaining vector
fields
$A_{\mu}^{I}$. The tensor fields $B_{\mu\nu}^{M}$ are inert under $K_S$.\\
\end{enumerate}
%
%
%
%
Turning on only the metric and the scalars, the corresponding
Lagrangian is of the form
\begin{equation}
e^{-1}\,\cL = -\frac{1}{2} R -\frac{1}{2}
(\partial_{\mu}\sigma)^2 -\frac{1}{2}g_{xy}\partial_{\mu}\varphi^x
\partial^{\mu}\varphi^y -V
\end{equation}
whereas the supersymmetry transformations of the fermions are
\begin{eqnarray}
\delta \psi_{\mu i} &=& D_{\mu} \epsilon_{i} -i\gamma_{\mu}
 \left(    g_{A} U_{i}{}^{j} + g_{S} S_{i}{}^{j}  \right) \epsilon_{j}
\label{N4gravitino}  \\
\delta \chi_{i}      &=& -\frac{i}{2} \prtsl \sigma \epsilon_{i} +3
\partial_{\sigma} \left(g_{A}  U_{i}{}^{j} + g_{S} S_{i}{}^{j}
\right) \epsilon_{j} \label{N4dilatino} \\
\delta \lambda_i^{a} &=& \frac{i}{2} f_{xi}^{a}{}^{j} ( \prtsl
 \varphi^{x} )     \epsilon_{j}                 
   -    \left( g_{A} V_{i}^{a}{}^{j} + g_{S}   T_{i}^{a}{}^{j}
 \right) \epsilon_{j} \label{N4gaugino}  \ .
\end{eqnarray}
Here, $g_{A}$ and $g_{S}$ are the gauge couplings of,
respectively, the Abelian and the non-Abelian gauge group, and
\begin{eqnarray}
U_{ij}&=& U_{ji}=
\frac{\sqrt2}{6}e^{2\sigma/\sqrt{3}}\Lambda^{N}_{M}L_{Nik}
L^{Mk}{}_{j} \label{Udef}\\
S_{ij} &=& S_{ji}= -\frac29 e^{-\sigma/\sqrt{3}} L^{J}_{ik} f_{JI}^K
L_{K}^{kl} L^{I}_{lj}, \label{Sdef}\\
V_{ij}^{a}&=& -V_{ji}^{a}=
\frac{1}{\sqrt{2}}e^{2\sigma/\sqrt{3}}\Lambda^{N}_{M}L_{Nij} L^{Ma} \label{Vdef}\\
T_{ij}^a &=&T_{ji}^{a}=  -e^{-\sigma/\sqrt{3}}
L^{Ja}L^K_{i}{}^{k}f_{JK}^I L_{Ikj}, \label{Tdef}
\end{eqnarray}
denote the ``fermionic shifts'' with the structure constants,
$f_{JI}^{K}$, of $K_{S}$ and the $K_{A}$ transformation matrix,
$\Lambda_{M}^{N}$, of the tensor fields $B_{\mu\nu}^{M}$.
 The
fermionic shifts also enter the scalar potential,
\begin{equation}
V=\frac12\,\Big[\; \ga^2\, V_{ij}^aV^{aij} -36\, \ga\gs\,
U_{ij}S^{ij} +\gs^2\,\Big(
 T_{ij}^aT^{aij}-9\, S_{ij}S^{ij}\Big)\,\Big],
\label{KAKS_pot}
\end{equation}
which is obtained from the trace of the ``Ward identity'' \cite{DHZ}
\begin{eqnarray}
\frac{1}{4} \delta_{i}^{j}\, V & = &     \frac12 \ga^2 V^a_{i}{}^k
V^a_{k}{}^{j}
 + \ga\gs\Big[9(S_i{}^k U_{k}{}^{j} +  U_i{}^k
    S_{k}{}^{j})
+ \frac12 (V^a_{i}{}^k T^a_{k}{}^{j}-T^a_i{}^k V^a_{k}{}^{j})\Big]\nnu\\
&&    -\frac12 \gs^2\,\Big[ T^a_i{}^k T^a_{k}{}^{j} - 9 S_i{}^k
S_{k}{}^{j}\Big]. \label{Wardidentity4}
\end{eqnarray}
%



\subsection{BPS-domain walls}
Our goal is to study domain walls of the form (\ref{curvedmetric})
that are supported by nontrivial scalar profiles $\sigma(r)$ and/or
$\varphi^{x}(r)$ and preserve one-half of the $\mathcal{N}=4$
supersymmetry. This analysis would be considerably simplified if one
could bring the BPS-equations and the scalar potential into a ``fake
supergravity''  form similar to (\ref{Freedman0a}) and
(\ref{Freedman1}) -- (\ref{Freedman3}) for the $\mathcal{N}=2$ case.
In $\mathcal{N}=2$ supergravity, the gravitino shift
$\mathbf{W}=-(ig/\sqrt{6})\mathbf{P}$ was a $\mathfrak{usp}(2) \cong
\mathfrak{su}(2)$-valued $(2 \times 2)$-matrix  (cf. eq.
(\ref{WP})). For $\mathcal{N}=4$ supergravity, the gravitino shift
is a $\mathfrak{usp}(4) \cong \mathfrak{so}(5)$-valued    $(4 \times
4)$-matrix,  $-i(g_{A} U_{i}{}^{j} +g_{S}S_{i}{}^{j})$, (see eq.
(\ref{N4gravitino})). In analogy with the $\mathcal{N}=2$ case, we
choose to call this gravitino shift $W_{i}{}^{j}$:
\begin{equation}
W_i{}^{j}:=-i \left(  g_{A}U_{i}{}^{j} +g_{S}S_{i}{}^{j}    \right)
. \label{Wdefinition}
\end{equation}
Furthermore, we will, from now on, suppress the
$\mathfrak{usp}(4)$ indices $i,j =1, \ldots, 4$ by using boldface
expressions such as
\begin{equation}
\mathbf{W} = W_i{}^{j}, \qquad \mathbf{W}\mathbf{W} =
W_i{}^{j}W_j{}^{k} \qquad \textrm{ etc.}, \label{boldface}
\end{equation}
just as we did in Section 2 for the analogous $(2 \times
2)$-matrices. Note that, in this boldface notation,  the position
of the indices is always assumed to be of the form shown in
(\ref{boldface}), which differs for example by a minus sign from
expressions such as $W_{ij}W^{jk}$ due to the convention
(\ref{summationconvention}). In a domain wall background, the
gravitino and dilatino    BPS-equations (\ref{N4gravitino}) and
(\ref{N4dilatino}) then take the form
\begin{eqnarray}
\left[ \nabla_{m}^{AdS_{4}} +\gamma_{m} \left( \frac{1}{2}
\,U^{\prime}\gamma_{5}
+\mathbf{W} \right) \right] \epsilon &=& 0, \label{N41}\\[2mm]
\left[ D_{r}+\gamma_{5}\mathbf{W}\right]
\epsilon &=& 0, \label{N42}\\[2mm]
 \left[ \gamma_{5} \sigma^{\prime} - 6\partial_\sigma
 \mathbf{W}\right] \epsilon &=& 0, \label{N43}
\end{eqnarray}
which are precisely of the same form as the fake supergravity
equations (\ref{Freedman1}) -- (\ref{Freedman3}), except, that
$\mathbf{W}$ is now a  $(4 \times 4)$-matrix instead of a
      $(2 \times 2)$-matrix.
Adding $0=
6g_{A}^{2}\mathbf{U}^2-\frac{9}{2}g_{A}^{2}(\partial_{\sigma}\mathbf{U})^2$
to the  right hand side of (\ref{Wardidentity4}),  the scalar
potential reads
\begin{equation}
\frac{1}{4} V \unity_{4} = -6\mathbf{W}^{2}
+\frac{9}{2}(\partial_{\sigma} \mathbf{W})^2  +\frac{1}{2} \Big[
g_{A}^{2}\mathbf{V}^a \mathbf{V}^a +g_{A}g_{S}
[\mathbf{V}^a,\mathbf{T}^a ] -g_{S}^2 \mathbf{T}^a\mathbf{T}^a \Big]
\label{Wardidentity4a}.
\end{equation}
Thus, if the domain wall is supported by $\sigma(r)$ only (i.e., if
$\varphi^{x\prime}=0$ and hence $\mathbf{V}^a=\mathbf{T}^a=0$ via eq.
(\ref{N4gaugino})), the BPS-equations (\ref{N41}) -- (\ref{N43})
and the scalar potential  (\ref{Wardidentity4a}) generalize
the $\mathcal{N}=2$ fake supergravity equations (\ref{Freedman0}) --
(\ref{Freedman3}) to what one might call ``$\mathcal{N}=4$ fake supergravity''.
The interesting question now is: Can a non-trivial profile
$\varphi^{x}(r)$ also be incorporated in this formalism?
This obviously requires two things:
\begin{enumerate}
\item The gaugino/tensorino  BPS-condition  (\ref{N4gaugino})
   has to be brought into
a form in which $\mathbf{V}^{a}$ and $\mathbf{T}^{a}$ are expressed in terms of derivatives of $\mathbf{W}$ with respect to $\varphi^{x}$ with the same relative prefactors  as in (\ref{N43}).
\item The term in brackets  in the scalar potential (\ref{Wardidentity4a}) should likewise
be re-expressed in terms of $\varphi^{x}$-derivatives of  $\mathbf{W}$
with the prefactor $9/2$, just as for the $(\partial_{\sigma} \mathbf{W})^{2}$-term.
\end{enumerate}
As we will  see, the  rewriting of   the vector- and tensor multiplet sector
along these lines  bears  many similarities with   the hypermultiplet sector
of $\mathcal{N}=2$ supergravity, but also shows some novel features.
Let us start with the BPS-equation (\ref{N4gaugino}). In the domain wall background (\ref{curvedmetric})  it reads, after using  a 
projector of the form   (\ref{projepsilon}) (now with $(4\times 4)$-matrices),
\begin{equation}
-\frac{i}{2}\varphi^{x\prime}\mathbf{f}^a_x\mathbf{\Theta} -g_{A}\mathbf{V}^a-g_S \mathbf{T}^a=0.
\end{equation}
Multiplying by $\mathbf{f}_{y}^a$ from the left and using       \cite{Awada,DHZ}
\begin{equation}
\mathbf{f}_{y}^{a} \mathbf{f}_{x}^{a} =g_{yx}\unity_{4} + 4\mathbf{R}_{yx}, \label{ffgR}
\end{equation}
this becomes
\begin{equation}
-\frac{i}{2} \varphi^{x\prime} g_{yx}\mathbf{\Theta} - 2i \varphi^{x \prime}
\mathbf{R}_{yx}\mathbf{\Theta} -\mathbf{f}^{a}_{y}(g_{A}\mathbf{V}^{a} +g_{S}\mathbf{T}^{a}) =0. \label{Zwischen}
\end{equation}
Decomposing (\ref{Zwischen}) into symmetric and antisymmetric part, one obtains
\begin{eqnarray}
-\frac{i}{2} \varphi^{x\prime} g_{yx}\mathbf{\Theta} - i \varphi^{x \prime}
[\mathbf{R}_{yx},\mathbf{\Theta}] -\frac{g_{A}}{2}[\mathbf{f}^{a}_{y},
\mathbf{V}^{a}] -  \frac{g_{S}}{2}\{ \mathbf{f}^{a}_{y}, \mathbf{T}^{a} \}  &=&0\label{newBPS1}\\
 - i \varphi^{x \prime}
\{ \mathbf{R}_{yx},\mathbf{\Theta} \} -\frac{g_{A}}{2} \{  \mathbf{f}^{a}_{y},
\mathbf{V}^{a}  \}   -  \frac{g_{S}}{2} [ \mathbf{f}^{a}_{y}, \mathbf{T}^{a}  ]  &=&0\label{newBPS2}.
\end{eqnarray}
Using (\ref{diff1}) -- (\ref{dC}) and the invariance conditions for the
structure constants $f_{IJ}^{K}$ and the transformation matrices
$\Lambda_{M}^{N}$,
\begin{eqnarray}
C_{IJ}f^{I}_{KL}+C_{IL}f^{I}_{KJ}&=&0\\
\Lambda_{M}^{P}C_{PN}+\Lambda_{N}^{P}C_{MP}&=&0,
\end{eqnarray}
one derives \cite{DHZ} 
\begin{eqnarray}
D_{y}  \mathbf{U}     &=&\frac{1}{6} [ \mathbf{f}^{a}_{y},\mathbf{V}^{a} ]\label{DyU}\\
D_{y} \mathbf{S}  &=& \frac{1}{6} \{ \mathbf{f}^{a}_{y}, \mathbf{T}^{a} \} \label{DyS}   ,
\end{eqnarray}
so that   (\ref{newBPS1}) becomes
\begin{equation}
\varphi^{x\prime}g_{yx} \mathbf{\Theta} +2 \varphi^{x\prime}     [  \mathbf{R}_{yx}, \mathbf{\Theta} ] +6 D_{y} \mathbf{W}=0. \label{newBPS3}
\end{equation}
If one now switches to adapted coordinates $(\varphi(r), \varphi^{\hat{x}})$,
with $\varphi^{\hat{x}}$ constant and perpendicular to $\varphi$ along a given
 flow curve, one obtains for the (canonically normalized)
  $\varphi$-component of (\ref{newBPS3})
\begin{equation}
\varphi^{\prime}\mathbf{\Theta} +6 D_{\varphi}\mathbf{W}=0.  
\label{gauginofake}
\end{equation}
Gauging away the composite  $Usp(4)$ connection,
\begin{equation}
\varphi^{x\prime} Q_{xi}{}^{j}=0 \qquad (Usp(4)\textrm{-gauge choice)}    \label{Usp4gaugechoice}
\end{equation}
and taking into account (\ref{projepsilon}),
this assumes the desired fake supergravity form (cf. (\ref{N43})). The other BPS-equations can again be viewed as constraints that determine the position of the flow curve on   the full scalar manifold.
Let us now turn to the scalar potential. Using (\ref{DyU}) and (\ref{DyS})
as well as (\ref{finverse}), one derives
\begin{eqnarray}
D_x\mathbf{U} D^{x} \mathbf{U} &=&-\frac{4}{9} \mathbf{V}^a \mathbf{V}^a
\label{VV}\\
\{ D_{x} \mathbf{S} ,  D^x \mathbf{U} \} &=& \frac{1}{3}[ \mathbf{T}^a, \mathbf{V}^a ] \label{VT}\\
D_x \mathbf{S} D^{x} \mathbf{S} &=& \frac{1}{9} \Big[ \mathbf{T}^a
\mathbf{T}^a +\frac{1}{2} \textrm{Tr} (\mathbf{T}^a \mathbf{T}^a ) \unity_{4} \Big]\label{TT}.
\end{eqnarray}
These relations allow one to re-express
the term in brackets    in         (\ref{Wardidentity4a})
in terms of derivatives of $\mathbf{U}$ and $\mathbf{S}$:
\begin{eqnarray}
\frac{1}{4} V \unity_{4}
&=& -6\mathbf{W}^{2}
+\frac{9}{2}(\partial_{\sigma} \mathbf{W})^2
 -\frac{9}{8} g_{A}^{2} D_{x}\mathbf{U} D^{x}\mathbf{U}-\frac{3}{2}
g_{A}g_{S} \{ D_{x}\mathbf{S}, D^{x}  \mathbf{U} \}\nonumber\\
& &  -\frac{9}{2}g_{S}^{2}
\Big( D_x\mathbf{S} D^{x} \mathbf{S}-\frac{1}{6} \textrm{Tr} (D_{x}\mathbf{S}
D^{x} \mathbf{S} ) \unity_{4} \Big) \label{VWUS}
\end{eqnarray}
We note three interesting  features of   this expression:
\begin{enumerate}
\item In contrast to the $\mathcal{N}=2$ analogue (\ref{VPP2}),
the Ward identity (\ref{VWUS}) can, in general, not be written without taking some traces.  
\item Even after taking the trace of (\ref{VWUS}), the prefactor
of the $(D_x\mathbf{U})^2$-term is different
from the prefactor  of the $\{ D_x\mathbf{S}, D^x \mathbf{U} \} $-term
and the $(D_x\mathbf{S})^2$-term. This means that, as long as
$g_{A}$ and $g_{S}$ are both non-vanishing,
   one    cannot write these terms
as something proportional to $(D_x\mathbf{W})^2$, i.e., in terms of derivatives of the \emph{full} gravitino shift $\mathbf{W}$. Again, this is different from the $\mathcal{N}=2$ case (\ref{VPP2}).
\item  If $g_{S}=0$, or if $g_{A}=0$, the scalar potential \emph{can}
be written as the full  gravitino shift and its derivatives, but
in none of these two special cases, the $\varphi^{x}$-derivatives
appear with the ``right''  cofficient $9/2$ required by fake supergravity.
 \end{enumerate}
Properties (i) and (ii) are clearly different from the $\mathcal{N}=2$
case. 
These differences
can in part be traced  to the fact that 
the adjoint  of $Usp(4)$ is no longer equivalent to 
the vector representation of
$SO(5)$, as was the case for $SU(2)$ and $SO(3)$. This implies, in particular,
that symmetric products of $\mathfrak{usp}(4)$-valued matrices such as $D_{x}\mathbf{S}D^{x}\mathbf{S}$
 are no 
longer automatically 
proportional to the unit matrix, as was the case for   
$\mathfrak{su}(2)$-valued matrices  such as $D_{X}\mathbf{P} D^{X} \mathbf{P}$      in the $\mathcal{N}=2$ case 
  due to the anticommutation properties of the Pauli-matrices, i.e., the Clifford algebra of $SO(3)$.
On the other hand, even
the $\mathcal{N}=2$ hypermultiplet sector did not fall into the $\mathcal{N}=2$
fake supergravity framework before   the BPS-equations and adapted coordinates were imposed (see eqs. (\ref{VPP2}) vs. (\ref{VPPsimple2})). Thus, there is still some hope that imposing the BPS-equations (\ref{newBPS1}) -- (\ref{newBPS2})
and using the  adapted coordinates $\varphi^{x}(r)=(\varphi(r),\varphi^{\hat{x}})$ miraculously transforms the last three terms in (\ref{VWUS}) into
$9/2 (D_{\varphi} \mathbf{W})^2$ and removes    at least some of the
above-mentioned   differences to the $\mathcal{N}=2$ case. 
To see whether this works, let us go back to the  BPS-condition (\ref{newBPS3}) and its $\varphi$-component
(\ref{gauginofake}), which we rearrange as (normalizing $g_{\varphi\varphi}=1$)
\begin{eqnarray}
D_{y}\mathbf{W}&=&-\frac{1}{6}\varphi^{x\prime}g_{xy}\mathbf{\Theta}-\frac{1}{3}\varphi^{x\prime} [ \mathbf{R}_{yx},\mathbf{\Theta} ] \label{trick1}\\ 
\Longrightarrow   D_{\varphi} \mathbf{W} &=& -\frac{1}{6} \varphi^{\prime} \mathbf{\Theta} \qquad (y=\varphi)
\label{trick2}
\end{eqnarray}
Squaring (\ref{trick2}) gives
\begin{equation}
D_{\varphi}\mathbf{W} D_{\varphi} \mathbf{W} = \frac{1}{36} (\varphi^{\prime})^2
\unity_{4} \label{trick3}
\end{equation}
On the other hand, squaring (\ref{trick1}) and using (\ref{newBPS2}) as well as
the identity
\begin{equation}
\mathbf{R}_{xy} \mathbf{R}^{x}{}_{z}= -\frac{1}{4} g_{yz} \unity_{4}
-\frac{3}{4} \mathbf{R}_{yz},
\end{equation}
one derives
\begin{eqnarray}
D_{y} \mathbf{W} D^{y}\mathbf{W} &=& \frac{5}{36} \varphi^{x\prime} \varphi^{z\prime} g_{xz} \unity_{4} -\frac{g_{A}^2}{36} \{\mathbf{f}_{y}^{a},\mathbf{V}^a \} \{ \mathbf{f}^{by}, \mathbf{V}^b \} \nonumber \\
& & -\frac{g_{A}g_{S}}{36} \Big( \{ \mathbf{f}_{y}^{a},\mathbf{V}^a \} [ \mathbf{f}^{by}, \mathbf{T}^b ] + [\mathbf{f}_{y}^{a},\mathbf{T}^a ] \{ \mathbf{f}^{by}, \mathbf{V}^b \} \Big) \nonumber\\
& & -\frac{g_{S}^2}{36}  [ \mathbf{f}_{y}^{a},\mathbf{T}^a ] [ \mathbf{f}^{by}, \mathbf{T}^b ]. \label{trick4}
\end{eqnarray}
Using (\ref{finverse}), the vielbeins can be eliminated, and (\ref{trick4})
becomes
\begin{equation}
D_{y} \mathbf{W} D^{y}\mathbf{W} = \frac{5}{36} \varphi^{x\prime} \varphi^{z\prime} g_{xz} \unity_{4} -\frac{g_{A}^{2}}{9} \mathbf{V}^a\mathbf{V}^a 
  -\frac{g_{A}g_{S}}{9} [ \mathbf{V}^{a}, \mathbf{T}^{a} ]
+\frac{g_{S}^{2}}{18} \textrm{Tr}(\mathbf{T}^{a} \mathbf{T}^{a} ) \unity_{4}.
\label{trick5}
\end{equation}
However,  $(D_x \mathbf{W})^2$ can be computed directly from (\ref{VV}) -- (\ref{TT}) and the definition (\ref{Wdefinition}):
\begin{equation}
D_{y} \mathbf{W} D^{y}\mathbf{W} = \frac{4}{9} g_{A}^2 \mathbf{V}^a 
\mathbf{V}^a  -\frac{g_{A}g_{S}}{3} [\mathbf{T}^a, \mathbf{V}^a ] 
 -\frac{g_{S}^2}{9} \Big(   \mathbf{T}^a \mathbf{T}^a    + \frac{1}{2} \textrm{Tr}(\mathbf{T}^a \mathbf{T}^a) \unity_{4} \Big) . 
  \label{trick6}
\end{equation}
Consistency of (\ref{trick5}) and (\ref{trick6}) then implies
\begin{equation}
\frac{5}{4}\varphi^{x\prime}\varphi^{z\prime}g_{xz} \unity_{4}
= 5g_{A}^{2} \mathbf{V}^{a} \mathbf{V}^{a} +4g_{A}g_{S} [ \mathbf{V}^{a},\mathbf{T}^a ]    -g_{S}^{2} \Big(  \mathbf{T}^{a} \mathbf{T}^{a} +\textrm{Tr}(\mathbf{T}^{a}\mathbf{T}^{a} ) \unity_{4}   \Big)    \label{trick7},
\end{equation}  
or, after taking the trace,
\begin{equation}\label{trick8}
\varphi^{x\prime}\varphi^{z\prime}g_{xz}= g_{A}^{2} \textrm{Tr}(\mathbf{V}^a
\mathbf{V}^{a}) -g_{S}^{2} \textrm{Tr} (\mathbf{T}^{a}\mathbf{T}^{a})
\end{equation}
Switching to adapted coordinates, (\ref{trick3}) then becomes
\begin{equation}
D_{\varphi}\mathbf{W}D_{\varphi}\mathbf{W}= \frac{1}{36} \Big(g_{A}^{2}
\textrm{Tr} (\mathbf{V}^a
\mathbf{V}^{a}) -g_{S}^{2} \textrm{Tr} (\mathbf{T}^{a}\mathbf{T}^{a})  \Big)
\unity_{4},\end{equation}
and we finally obtain for the scalar potential (\ref{Wardidentity4a})
\begin{equation}
V=-6\textrm{Tr}\mathbf{W}^2 +\frac{9}{2}\textrm{Tr}(\partial_{\sigma}\mathbf{W})^2 +\frac{9}{2}\textrm{Tr}(D_{\varphi}\mathbf{W})^2. \label{trick10}
\end{equation}
Thus, after employing  the $Usp(4)$ gauge choice (\ref{Usp4gaugechoice}),
the $\varphi$-sector and the $\sigma$-sector enter the theory symmetrically
and with the              ``right'' prefactors. 
If both $\varphi$ and $\sigma$ are running,
 one can, just as   in $\mathcal{N}=2$ supergravity,
  go over to a total adapted coordinate $\phi(r)$ with
 $\sigma^{\prime}\partial_{\sigma} +\varphi^{\prime} \partial_{\varphi}
=\phi^{\prime}\partial_{\phi}$ so as to obtain  $\mathcal{N}=4$
  single-field
 fake supergravity equations
   (see the discussion around
(\ref{coordtrphi})).



\subsection{Consistency conditions and   domain wall curvature}
Let us summarize what we have shown so far. In 5D,
$\mathcal{N}=4$ supergravity,  the             gravitino BPS-equations    in a
$\frac{1}{2}$-supersymmetric domain wall background read
\begin{eqnarray}
\left[ \nabla_{m}^{AdS_{4}} +\gamma_{m} \left( \frac{1}{2}
\,U^{\prime}\gamma_{5}
+\mathbf{W} \right) \right] \epsilon &=& 0, \label{trick11}\\[2mm]
\left[ D_{r}+\gamma_{5}\mathbf{W}\right]
\epsilon &=& 0.  \label{trick12}
\end{eqnarray}
Subjecting the spinor $\epsilon$ to
\begin{equation}\label{trick12a}
\gamma_{5}\epsilon =-\mathbf{\Theta}\epsilon,
\end{equation}
the dilatino equation becomes
\begin{equation}
  \sigma^{\prime} \mathbf{\Theta}  + 6\partial_\sigma
 \mathbf{W} = 0, \label{trick13}
\end{equation}
and the gaugino/tensorino BPS-equation can be decomposed as follows
\begin{eqnarray}
\varphi^{x\prime} g_{yx}\mathbf{\Theta} +2 \varphi^{x \prime}
[\mathbf{R}_{yx},\mathbf{\Theta}] + 6 D_{y} \mathbf{W}  &=&0\label{trick14}\\[2mm]
 2 \varphi^{x \prime}
\{ \mathbf{R}_{yx},\mathbf{\Theta} \} -ig_{A} \{  \mathbf{f}^{a}_{y},
\mathbf{V}^{a}  \}   -  ig_{S} [ \mathbf{f}^{a}_{y}, \mathbf{T}^{a}  ]  &=&0\label{trick15}.
\end{eqnarray}
In adapted coordinates, $\varphi^{x}(r)=(\varphi(r),\varphi^{\hat{x}})$,
the scalar potential reads
\begin{equation}
V=-6\textrm{Tr}\mathbf{W}^2 +\frac{9}{2}\textrm{Tr}(\partial_{\sigma}\mathbf{W})^2 +\frac{9}{2}\textrm{Tr}(D_{\varphi}\mathbf{W})^2  ,  \label{trick16}
\end{equation}
and (\ref{trick14})    splits further into
\begin{eqnarray}
\varphi^{\prime}\mathbf{\Theta} +6D_{\varphi}\mathbf{W}&=&0\label{trick16a}\\
2 \varphi^{ \prime}
[\mathbf{R}_{\hat{y}\varphi},\mathbf{\Theta}] + 6 D_{\hat{y}} \mathbf{W}  &=&0 . 
\label{trick16b}
\end{eqnarray}
Eqs.   (\ref{trick13})    and (\ref{trick14}) can be solved for $\mathbf{\Theta}$:
\begin{eqnarray}
\mathbf{\Theta} &=& -6\frac{\partial_{\sigma}\mathbf{W}}{\sigma^{\prime}}\label{trick17}\\
\mathbf{\Theta} &=& -6 \frac{\varphi^{y\prime}D_{y}\mathbf{W}}{\varphi^{x\prime}\varphi^{z\prime}g_{xz}} = -6\frac{D_\varphi \mathbf{W}}{\varphi^{\prime}} . 
\label{trick18}
\end{eqnarray}
If both $\sigma$ and $\varphi^{x}$ are running, consistency of these two expressions requires
\begin{equation}
\frac{\partial_{\sigma}\mathbf{W}}{\sigma^{\prime}} = \frac{\varphi^{y\prime}D_{y}\mathbf{W}}{\varphi^{x\prime}\varphi^{z\prime}g_{xz}} =
\frac{D_\varphi             \mathbf{W}}{\varphi^{\prime}}.\label{trick19}
\end{equation}
Eqs. (\ref{trick17}) and (\ref{trick18})  also imply
\begin{equation}
D_{r}\mathbf{W}   \equiv        (     \sigma^{\prime}\partial_{\sigma} +  \varphi^{x\prime} D_{x})\mathbf{W} = -\frac{1}{6} ((\sigma^{\prime})^{2}
+\varphi^{x\prime}g_{xy}\varphi^{y\prime} ) \mathbf{\Theta}. \label{trick19a}
\end{equation}
Squaring (\ref{trick17}) and (\ref{trick18}) gives the first order equation for the scalars,
\begin{eqnarray}
\sigma^{\prime}&=&\pm 6 \sqrt{(\partial_{\sigma}\mathbf{W})^2}\label{trick20}\\
\varphi^{x\prime}g_{xy}\varphi^{y\prime}&=&\pm 6
\sqrt{(\varphi^{x\prime}D_{x}\mathbf{W})^2}    \Rightarrow \varphi^{\prime} = \pm 6 \sqrt{(D_{\varphi}\mathbf{W})^2}\label{trick21}
\end{eqnarray}
Let us now turn to the warp factor. Just as in section 2.2, the integrability condition of
(\ref{trick11}), yields
\begin{equation}
(U^{\prime})^2 = 4\mathbf{W}^{2} - \frac{e^{-2U}}{L_{4}^2}.\label{trick22}
\end{equation}
On the other hand, the compatibility condition between (\ref{trick11}) and (\ref{trick12a}) gives
\begin{equation}
U^{\prime}=\{ \mathbf{\Theta}, \mathbf{W} \} ,     \label{trick23}
\end{equation}
which, when combined with (\ref{trick22}), gives
\begin{equation}
-\frac{e^{-2U}}{L_{4}^2} \unity_{4}= \{\mathbf{\Theta},\mathbf{W} \}^2 -
4\mathbf{W}^{2}, \label{trick24}
\end{equation}
just as in (\ref{algebraic}). Hence, a  BPS-domain wall is flat if and only if
\begin{equation}
\{\mathbf{\Theta},\mathbf{W} \}^2=4\mathbf{W}^{2}   \qquad \textrm{(flatness condition)}.   \label{trick25}
\end{equation}
It is shown   in Appendix A that this is equivalent to
\begin{equation}
[ \mathbf{\Theta}, \mathbf{W} ] =0 \Leftrightarrow  [ D_{r}\mathbf{W} , \mathbf{W} ] =0 \qquad \textrm{(equivalent flatness condition)}, \label{trick26}
\end{equation}
where we have used (\ref{trick19a}).
Finally, there is the $\mathcal{N}=4$ analogue of eq. (\ref{commutator1}), namely the compatibility condition of (\ref{trick12}) and (\ref{trick12a}),
\begin{equation}
 [ \mathbf{\Theta} , D_{r} \mathbf{\Theta}- 2 \mathbf{W} ] =0  ,  \label{trick27}
\end{equation}
or, using (\ref{trick19a}),
\begin{equation}
\Big[ D_{r}\mathbf{W} , D_{r} D_{r} \mathbf{W} + \frac{1}{3} ((\sigma^{\prime})^{2}+\varphi^{x}\varphi^{y}g_{xy}) 
\mathbf{W}  \Big] =0. \label{trick28}
\end{equation}
%



 \subsection{Special cases}
In this subsection, we will take a closer look at the   implications
of the equations listed in section 3.4 and apply them to a number of special cases.\\
\vspace{-1mm}\\
\textbf{Case 1:} The gauging is purely Abelian:
\begin{equation}
g_{S}=0.
\end{equation}
In this case, $\mathbf{W}=-ig_{A}\mathbf{U}$, and (\ref{trick13}) implies
\begin{equation}
\sigma^{\prime} \mathbf{\Theta} = -4\sqrt{3} \mathbf{W}.\label{trick29}
\end{equation}
Thus, $\sigma^{\prime}=0$ would imply $\mathbf{W}=0$ along the flow, and, hence, because of (\ref{trick23}), also $U(r)= \textrm{const.}$, i.e., a trivial domain wall. In other words, for a non-trivial Abelian      BPS-domain wall,
$\sigma^{\prime}$ has to   be non-zero. Now, a                
   running $\sigma(r)$, however, means that (\ref{trick29}) can be solved for $\mathbf{\Theta}$:
\begin{equation}
\mathbf{\Theta}= -4\sqrt{3}\frac{\mathbf{W}}{\sigma^{\prime}}.
\end{equation}
This implies that the domain wall has to be \emph{flat}
 because of (\ref{trick26}).
On the other hand, due to the simple $\sigma$-dependence,      (\ref{trick20})
can be readily solved:
\begin{equation}
\sigma(r)= a \ln (r-r_{o}) +b
\end{equation}
with some constants $a$ and $b$. Hence, $\sigma(r)$ always approaches infinity,
which is not surprising in view of the simple runaway behaviour of the scalar potential in the $\sigma$-direction  when $g_{S}=0$.\\
\vspace{-1mm}\\
\textbf{Case 2:} The gauging is purely Non-Abelian:
\begin{equation}
g_{A}=0.
\end{equation}
In this case, $\mathbf{W}=-i g_{S} \mathbf{S}$, and the simple exponential behaviour of $\mathbf{W}$   leads to similar conclusions as in the purely Abelian case: Just as in (\ref{trick29}), one concludes that $\sigma(r)$ has to have
a non-trivial $r$-dependence for the domain wall to be non-trivial.
This, however,                       
 also implies that $\mathbf{\Theta}$ is always proportional
to $\mathbf{W}$, and any domain wall must be flat, due to (\ref{trick26}).
Again, the runaway behavior of the potential leads to a logarithmic
$r$-dependence of the scalar field $\sigma(r)$.\\
\vspace{-1mm}\\
\textbf{Case 3:} The mixed gauging:
\begin{equation}
g_{A}g_{S}\neq 0.
\end{equation}
When both the Abelian part and the non-Abelian part are gauged, the
$\sigma$-dependence of $\mathbf{W}$ no longer factors out and neither does the scalar potential have a simple   runaway behaviour in the $\sigma$-direction.
Nevertheless, the structure of domain walls that are solely supported by the supergravity scalar $\sigma$ are still  quite limited. In order to see this, consider the consistency condition (\ref{trick28}), which, for
$\varphi^{x\prime}=0$, simplifies to
\begin{equation}
 \Big[ \partial_{\sigma} \mathbf{W} , \partial_{\sigma}^{2} \mathbf{W} + \frac{1}{3} \mathbf{W}  \Big] =0. \label{trick30}
\end{equation}
Using (\ref{Wdefinition}) and the $\sigma$-dependence of $\mathbf{U}$ and $\mathbf{S}$,   (\ref{Udef}) -- (\ref{Sdef}), one easily sees that  (\ref{trick30})
implies
\begin{equation}
[\mathbf{U},\mathbf{S}]=0. \label{trick31}
\end{equation}
But this also implies $[\partial_{\sigma}\mathbf{W},\mathbf{W}]=0$, and hence,
 via (\ref{trick26}), the flatness of the domain wall. Thus, taking into account  our observations in the purely Abelian and purely non-Abelian case,
we find that a domain wall supported by $\sigma(r)$ only can never be curved.
Obviously, the condition $[\mathbf{U},\mathbf{S}]=0$ is automatically satisfied if either $g_{A}$ or $g_{S}$ vanishes. One way to satisfy $[\mathbf{U},\mathbf{S}]=0$ for $g_{A}g_{S}\neq 0$ is as follows:  Suppose, the gauge group
$K$
contains the obvious  $SO(3)\times SO(2)$ subgroup of $SO(5)\subset SO(5,n)$
as a factor. More generally, one could take the $SO(2)$ factor of the gauge group
 to be a
diagonal subgroup of the $SO(2)\subset SO(5)$ and some product of $SO(2)$'s
that are contained in $SO(n)\subset SO(5,n)$ such that, under $K_{A}$-transformations,
the tensors charged under $SO(2)\subset SO(5)$ do not mix with the tensors charged under the
other $SO(2)$-subgroups of $SO(n)$. These gaugings are precisely the ones that
occur in the $\mathcal{N}=4$ orbifold compactifications \cite{Kachru:1998ys}
in the AdS/CFT-correspondence \cite{CGWZ}. In such a gauging, the five   vector fields
$A_{\mu}^{1,\ldots,5}$
of the ungauged   supergravity multiplet split into a triplet of $SO(3)$-gauge fields, which we take to be
$A_{\mu}^{1,2,3}$, and a doublet of tensor fields, $B_{\mu\nu}^{4,5}$, charged
under the $SO(2)$.
Suppose further that the scalar fields $\varphi^{x}$ of the vector- and tensor multiplets all sit at the  ``origin''   of the symmetric space $SO(5,n)/SO(5)\times SO(n)$: $L_{\tilde{I}}^{A}=\delta_{\tilde{I}}^{A}$. Since, by assumption, the tensor field transformation matrix
$\Lambda_{M}^{N}$ does not mix the supergravity tensors   $B_{\mu\nu}^{4,5}$
with the other tensor fields (provided such additional tensor fields exist), and since the $SO(3)$ part of the gauge group is supposed to be a direct factor, it is easy to see that $V_{ij}^{a}$ and $T_{ij}^{a}$ from eqs. (\ref{Vdef}) and (\ref{Tdef})
vanish at this critical point, which is consistent with the    BPS-conditions   (\ref{newBPS1}) -- (\ref{newBPS2}) and $\varphi^{x\prime}=0$. Using $SO(5)$-gamma matrices    (see eq. (\ref{SO5Gamma})) 
to convert $Usp(4)$-indices $i,j=1,\ldots,4$ into  $SO(5)$ indices
$\alpha,\beta=1,\ldots 5$, it is easy to see that
\begin{eqnarray}
\mathbf{U}&\sim& e^{\frac{2\sigma}{\sqrt{3}}}\mathbf{\Gamma}_{45}\\
\mathbf{S}&\sim& e^{-\frac{\sigma}{\sqrt{3}}} \mathbf{\Gamma}_{123} \sim
e^{-\frac{\sigma}{\sqrt{3}}}\mathbf{\Gamma}_{45}
\end{eqnarray}
and hence $[\mathbf{U},\mathbf{S}]=0$. In fact, if the scalars $\varphi^{x}$ are frozen as described, the model becomes effectively the Romans model \cite{Romans}.
The BPS-conditions for the general case of running
scalars $\varphi^{x}(r)$ and $\sigma(r)$ can be analyzed along similar lines using
the equations (\ref{DyU}), (\ref{DyS}) as well as \cite{DHZ}
\begin{eqnarray}
D_{x}\mathbf{V}^{a}&=&     \Big(    \frac{e^{\frac{2\sigma}{\sqrt{3}}}}{2\sqrt{2}}\Lambda_{M}^{N}L_{N}^{a}L^{Mb}          \Big)    \mathbf{f}_{x}^{b} +\frac{3}{2} [    \mathbf{f}_{x}^{a}, \mathbf{U} ]\\
D_{x}\mathbf{T}^{a} &=&    \Big(   \frac{e^{-\frac{\sigma}{\sqrt{3}}}}{2}f_{JK}^{I} L^{Ja}L_{I}^{b}     \Big)    [\mathbf{L}^{K} , \mathbf{f}_{x}^{b} ] + \frac{3}{2} \{  \mathbf{f}_{x}^{a}, \mathbf{S} \},
\end{eqnarray}
 but the results are in general model-dependent and beyond the scope of this paper.

\section{Conclusions}
In this paper, we have shown that the BPS-equations and the scalar potentials for
             $1/2$-BPS-domain walls in 5D, $\mathcal{N}=4$ gauged supergravity
can be cast into a generalized form of ``fake'' supergravity. In many respects, this parallels the situation in $\mathcal{N}=2$ supergravity in five dimensions, but there are also important differences. Most importantly, the gravitino shift is now a $\mathfrak{usp}(4)$-valued $(4\times 4)$-matrix instead of a
$\mathfrak{su}(2)$-valued    $(2 \times 2)$-matrix. This means that some
peculiarities of the group $SU(2)$ are no longer valid, which makes it all the more surprising that the simple form of the fake supergravity equations remains largely unaltered (in fact, the only immediately visible difference is that the scalar potential in $\mathcal{N}=4$ fake supergravity can no longer be written in terms of            the gravitino shift                           $\mathbf{W}$ without taking the trace).
Furthermore, due to the doubled amount of supersymmetry, there are now twice as many   BPS-conditions to satisfy. It should also be noted that the scalar manifolds are no longer of the type encountered in    $\mathcal{N}=2$ supergravity,   but
are subject to completely different         geometrical constraints.
The fact that nevertheless the full dynamics along the flow line is
captured by almost identical equations suggests  
 that possibly   all BPS-domain walls in all spacetime dimensions and for all amounts of supersymmetry   can be described in terms of an appropriately generalized form of fake supergravity. This could, for instance, be due to some general
properties of gauged supergravities in the spirit  of   \cite{Ferrara}.   The results of this paper should help distinguish $\mathcal{N}=2$ artifacts from  this   general formulation.

Recasting  domain wall equations of supergravity theories into a ``fake'' supergravity language greately simplifies their study. This was already demonstrated in \cite{CCDVZ}, and in this paper we saw that also $\mathcal{N}=4$ domain walls can be studied  quite efficiently   in this language. For example, we could easily rule out curved BPS-domain walls if the gauge group is purely Abelian or purely semisimple.
In both of these cases,           domain walls   furthermore    show    a runaway behavior in the $\sigma$-direction. In the mixed gauging, curved domain walls are ruled out if they are supported by $\sigma(r)$ only.

It might be interesting to apply the results of this paper to the study of 
            holographic    renormalization group flows along the lines of \cite{CGWZ} or in the context of       domain wall solutions in  
 flux compactifications, which single out certain types of gauge groups
         \cite{Mayer:2004sd,      House:2004hv,       LouisVaula}. Another interesting further direction
concerns a generalization to    $1/4$-BPS-domain walls \cite{CGWZ}, 
which are impossible in $\mathcal{N}=2$ supergravity.
 One   might  also  wonder whether other types of solutions  such as charged black holes or cosmic strings have a similar description in terms of some other form of  
 ``fake'' supergravity, in which all scalars are treated equally, independently of the
                            spacetime dimension or the amount of supersymmetry.


\begin{appendix}
\section{The flatness condition}
As  shown in section 3.3, a $1/2$-BPS domain wall       in 5D, $\mathcal{N}=4$ supergravity   is flat if and only if     (cf. eq. (\ref{trick25})) 
\begin{equation}
\{ \mathbf{\Theta},\mathbf{W} \}^{2}=4 \mathbf{W}^{2}. 
\label{Ausgangsgleichung}
\end{equation}
A sufficient condition for this to hold is obviously 
\begin{equation}
[\mathbf{\Theta},\mathbf{W} ] = 0. \label{neueGleichung}
\end{equation}
We will now show that this condition is also necessary.
Both $\mathbf{\Theta}$ and $\mathbf{W}$ are $\mathfrak{usp}(4)\cong \mathfrak{so}(5)$-valued, so they can be expressed in terms of the $SO(5)$ gamma matrices
(\ref{SO5Gamma}) via
\begin{equation}
\mathbf{\Theta}=\Theta^{\alpha\beta}\mathbf{\Gamma}_{\alpha\beta}, \qquad
\mathbf{W}=W^{\alpha\beta}\mathbf{\Gamma}_{\alpha\beta}.
\end{equation} 
Without loss of generality, we can assume $\mathbf{\Theta}=2\Theta^{12}\mathbf{\Gamma}_{12}$. Using
\begin{equation}
\{ \mathbf{\Gamma}_{\alpha\beta} , \mathbf{\Gamma}_{\gamma\delta} \} 
= 2\mathbf{\Gamma}_{\alpha\beta\gamma\delta} +2\delta_{\alpha\delta}\delta_{\beta\gamma} -2\delta_{\beta\delta}\delta_{\alpha\gamma},\label{gammarelation}
\end{equation}
one finds that $\mathbf{\Theta}^2=\unity_{4}$ implies $(\Theta^{12})^{2}=-1/4$.
(\ref{gammarelation}) now implies
\begin{equation}
\{ \mathbf{\Theta}, \mathbf{W} \}^{2} = [2\Theta^{\alpha\beta}W^{\gamma\delta}
\mathbf{\Gamma}_{\alpha\beta\gamma\delta} -4\Theta^{\alpha\beta}W^{\alpha\beta}\unity_{4} ]^2 \label{gammarelation2}.
\end{equation}
Isolating the part of (\ref{gammarelation2}) that is proportional to the unit matrix,   one easily sees that this is
    (remembering   $\mathbf{\Theta}=2\Theta^{12}\mathbf{\Gamma}_{12}$
and $(\Theta^{12})^{2}=-1/4$)
\begin{equation}
-16[(W^{12})^2 + (W^{34})^2 + (W^{35})^2 + (W^{45})^2 ].
\end{equation}
In $4\mathbf{W}^2$, on the other hand, the part proportional to the 
unit matrix is easily seen to be $-8W^{\alpha\beta}W^{\alpha\beta}$,
so that (\ref{Ausgangsgleichung}) implies that all components
except $W^{12},W^{34},W^{45},W^{35}$ have to vanish.
This implies (\ref{neueGleichung}).
\end{appendix}
\\
\vspace{5mm}\\
\textbf{Acknowledgements:} I have  benefitted from earlier discussions 
with A.~Celi, A.~Ceresole, G.~Dall'Agata, C.~Herrmann  and A.~Van Proeyen
 on the material of refs. \cite{CCDVZ,DHZ}. 
This work  is supported 
  by the   German Research Foundation (DFG)      within the 
Emmy Noether Program            (ZA 279/1-1).

\providecommand{\href}[2]{#2}\begingroup\raggedright\endgroup

\end{document}